\pdfoutput=1
\RequirePackage[colorlinks,allcolors=blue]{hyperref}
\documentclass[draft]{agujournal2019}
\usepackage{url} %this package should fix any errors with URLs in refs.
\usepackage{lineno}
\usepackage{soul}

\usepackage{amsmath}
\usepackage{hyperref}
\usepackage[nameinlink,noabbrev]{cleveref}
\usepackage{wrapfig}
\usepackage{siunitx}
\usepackage{graphicx}
\usepackage{mathtools}
\usepackage{hyphenat}
\usepackage{tikz}
\usepackage{ragged2e}
\justifying
\usetikzlibrary{positioning,shapes,arrows.meta,calc}

\newcommand\thefont{\expandafter\string\the\font}

\newcommand{\breakingcomma}{%
  \begingroup\lccode`~=`,
  \lowercase{\endgroup\expandafter\def\expandafter~\expandafter{~\penalty0 }}}

\DeclareSIUnit\month{month}
\DeclareSIUnit\UTZ{UTZ}

\usepackage{xr}
\makeatletter

\newcommand*{\addFileDependency}[1]{% argument=file name and extension
\typeout{(#1)}% latexmk will find this if $recorder=0
\@addtofilelist{#1}
\IfFileExists{#1}{}{\typeout{No file #1.}}
}\makeatother

\makeatletter
\newcommand{\dummylabel}[2]{\def\@currentlabel{#2}\label{#1}}
\makeatother

\draftfalse

\journalname{Journal of Advances in Modeling Earth Systems (JAMES)}

\begin{document}

\title{Interpretable multiscale Machine Learning-Based Parameterizations of Convection for ICON}

\authors{Helge Heuer\affil{1}, Mierk Schwabe\affil{1}, Pierre Gentine\affil{2}, Marco A. Giorgetta\affil{3}, Veronika Eyring\affil{1,4}}

\affiliation{1}{Deutsches Zentrum für Luft- und Raumfahrt e.V. (DLR), Institut für Physik der Atmosphäre, Oberpfaffenhofen, Germany}
\affiliation{2}{Center for Learning the Earth with Artificial Intelligence and Physics (LEAP), Columbia University, New York, NY, USA}
\affiliation{3}{Max Planck Institute for Meteorology, Hamburg, Germany}
\affiliation{4}{University of Bremen, Institute of Environmental Physics (IUP), Bremen, Germany}

\correspondingauthor{H. Heuer}{helge.heuer@dlr.de}

\begin{keypoints}
\item We train/benchmark machine learning models on convective fluxes derived from realistic coarse-grained data of storm-resolving simulations
\item Shapley values reveal that the best offline model, a U-Net, learns non-causal links to precipitation and shows poor online performance
\item A model, without non-causal precipitation connections, runs more stable coupled to ICON and indicates better precipitation predictions
\end{keypoints}

\begin{abstract}
Machine learning (ML)-based parameterizations have been developed for Earth System Models (ESMs) with the goal to better represent subgrid-scale processes or to accelerate computations.
ML-based parameterizations within hybrid ESMs have successfully learned subgrid-scale processes from short high-resolution simulations.
However, most studies used a particular ML method to parameterize the subgrid tendencies or fluxes originating from the compound effect of various small-scale processes (e.g., radiation, convection, gravity waves) in mostly idealized settings or from superparameterizations.
Here, we use a filtering technique to explicitly separate convection from these processes in simulations with the Icosahedral Non-hydrostatic modelling framework (ICON) in a realistic setting and benchmark various ML algorithms against each other offline.
We discover that an unablated U-Net, while showing the best offline performance, learns reverse causal relations between convective precipitation and subgrid fluxes.
While we were able to connect the learned relations of the U-Net to physical processes this was not possible for the non-deep learning-based Gradient Boosted Trees.
The ML algorithms are then coupled online to the host ICON model.
Our best online performing model, an ablated U-Net excluding precipitating tracer species, indicates higher agreement for simulated precipitation extremes and mean with the high-resolution simulation compared to the traditional scheme.
However, a smoothing bias is introduced both in water vapor path and mean precipitation.
Online, the ablated U-Net significantly improves stability compared to the non-ablated U-Net and runs stable for the full simulation period of 180 days.
Our results hint to the potential to significantly reduce systematic errors with hybrid ESMs.
\end{abstract}

\section*{Plain Language Summary}
Due to their computational costs, it is currently not feasible to run more accurate high-resolution climate models on a global domain on climate (century) time-scales.
However, high-accuracy climate simulations are needed for more robust and detailed projections of our future climate.
Here, we develop and evaluate various machine learning-based convection parameterizations learned on reconstructed and coarse-grained high-resolution subgrid fluxes to solve this problem, and benchmark their performance.
The data set is chosen from simulations of the Icosahedral Non-hydrostatic modelling framework (ICON) in a realistic setting of the tropical Atlantic and at storm-resolving resolutions.
We focus only on convective subgrid fluxes that are isolated from other components.
We improve the best ML algorithms further by excluding variables that cause unphysical correlations.
Finally, we explain the learned relations of the best data-driven schemes based on physical process understanding, test their performance when coupled to the ICON model, and achieve stable coupled simulations for 180 days as well as improved precipitation predictions.

\section{Introduction}
General Circulation Models (GCMs) have been used since the late 1960s to answer scientific questions about our climate \cite{RN156,RN155} and to project its expected changes, which are already felt across the globe \cite{Eyring2021}.
Over time, these models gradually included more and more aspects and processes of the climate system and have evolved into Earth System Models (ESMs), including the carbon cycle and biogeochemical processes.
However, the uncertainty of the simulated equilibrium climate sensitivity (ECS), i.e. the response of global surface air temperature to a doubling of CO2 at equilibrium, has not reduced significantly in the last decades \cite{RN169}.
For the latest generation of ESMs, the ECS is estimated by the Intergovernmental Panel on Climate Change (IPCC) Sixth Assessment Report \cite{AR6WG1CH7} at \qtyrange{2}{5}{\celsius}.
This uncertainty is about twice the uncertainty for the estimated ECS including all other scientific evidence such as emergent constraints and paleoclimates of \qtyrange{2.5}{4}{\celsius} \cite{AR6WG1CH7}.

A large portion of this uncertainty is attributed to cloud feedbacks \cite{Schneider2017,RN159}, the change in cloud types and distributions in response to warming climate.
Therefore, it is highly important to have a good representation of the effects of convection, which is typically a subgrid-scale process in climate models \cite{Sherwood2014}.
Parameterizations based on physical process understanding, normally relying on mass-flux approaches \cite{RN59,RN44}, have been used extensively for approximating the effect of subgrid convection on the large scale.
These parameterizations, however, cause some common problems in climate models \cite{Eyring2021a}, such as biases in precipitation patterns \cite{RN158,RN165,Fosser2024}, in the position and shape of the intertropical convergence zone (ITCZ) \cite{RN82}, the missing representation of convectively coupled waves, and the Madden-Julian Oscillation \cite{RN110}, or teleconnections \cite{Mahajan2023} and the incorrect diurnal cycle of convection \cite{doi:10.1073/pnas.1505077112}. These biases are reduced in storm-resolving models \cite{RN82,RN57,Bock2020,stevens2020added}.

Accurately representing convection in climate models remains a challenge due to its complex and multiscale nature.
In light of recent advances in deep learning, many data-driven machine learning-based parameterizations have been developed to reduce the above-mentioned biases \cite{RN127,RN5,RN19,RN107,iglesias-suarez_causal_nns,zanna2023arxiv}.
These studies first used multilayer perceptron (MLP) neural networks in a simplified aquaplanet setup to replace the superparameterized physics in the SuperParameterized Community Atmosphere Model (SPCAM3) \cite{CAM3}.
Random Forests (RFs) have been used as well \cite{RN2,RN17} with the advantage of guaranteeing conservation properties and physical consistency, via constraints in the sign of quantities such as precipitation, as well as on its magnitude (reducing coupled model instability).
A disadvantage of RFs is however that they do not extrapolate outside their training domain at all and so are inherently limited in their application for a changing climate.
They can also struggle to represent the diversity of complex data.

To combine conservation properties that are essential for a climate model, and the ability to extrapolate to some extent, \citeA{RN25} used MLPs to predict vertical fluxes instead of tendencies (the vertical convergence of the fluxes).
More recently, they extended their work by including convective momentum transport in an idealized aquaplanet setting as well \cite{RN145}.
\citeA{RN125} used residual neural networks to emulate the physical tendencies resulting from a superparameterization of moist physics and radiation in a realistic setting with coupled simulations running stably over 10 years.

With this work we build on previous studies on data-driven convection parameterizations and ML-based schemes, targeting the ICON model \cite{RN24,2023arXiv230408063G}.
We extend these approaches in several aspects.
We use high-resolution data that explicitly resolve convection and employ a coarse-graining method to calculate and isolate the convective mesoscale flux that is subgrid for a coarse climate model, here ICON in a real-world setting.
We benchmark a set of different machine learning methods trained on a realistic data set with orography (Dataset section).
Although it can be argued to what extent explicit process separation is sensible \cite{RN154}, most parameterization schemes act independently (in parallel or sequentially) from each other for different subgrid processes \cite{RN4}.
For this reason, simplicity, and because the trained ML models should be easily interoperable with the GCM in a coupled mode we treat convection as a separated process.
Furthermore, this enables us to use explainable Artificial Intelligence (AI) methods to interpret the ML models with respect to our physical understanding of atmospheric convection.
To focus on the effects of subgrid convection for coarse resolution simulations, where convection must be parameterized, we introduce a filtering technique to capture convective circulations as resolved in storm resolving simulations.
Apart from making it possible to selectively replace only the conventional parameterization, this approach allows to better interpret the physics of the learned ML model as it does not mix different processes such as convection and radiation.
We propose a new way of computing the coarse-grained target quantities by not neglecting horizontal fluctuations (not applying the Boussinesq approximation) in the density as is typical for Reynolds-averaging.
Additionally, we use an explainable AI technique to interpret the model predictions and relate the revealed connections to physical process understanding.
Similarly to the spectral analysis tool by \citeA{RN143}, this method builds trust in the retrieved models and can be used to evaluate the ML model, going beyond common metrics such as the root mean squared error (RMSE) or the coefficient of determination.

In the end we will test the stability of the U-Net when coupled to the ICON model.
Here we test the extrapolation capabilities of the ML models as they are trained on regional data and then applied on larger/global domains.

This paper is structured as follows.
First, in \cref{sec:data_and_preprocessing} we describe the data, preprocessing, and coarse-graining method.
Afterwards, we introduce the machine learning methods in \cref{sec:ml_methods}.
Results of the offline evaluation/benchmarking of different machine learning models are then shown and their predictions interpreted using an explainable AI technique in \cref{sec:results}.
We will conclude \cref{sec:results} with an online stability test of the developed U-Net parameterizations.
Finally, we discuss our results and give a conclusion of our work.

\section{Data and Preprocessing} \label{sec:data_and_preprocessing}
As training data we use short storm-resolving simulations of the tropical Atlantic that accompanied the NARVAL expeditions performed with ICON \cite{RN57,RN58}.
Focusing on the deep convective systems of the ITCZ and the explicit representation of convection, this data set serves as an ideal starting point to learn convective subgrid processes.
There were two related research campaigns, one from the boreal winter (Dec $2013$ / Jan $2014$), and one from the boreal summer (Aug $2016$).
We use simulation data accompanying both expeditions.
The horizontal resolution of the used simulations is $\Delta x \approx \qty{2.5}{\kilo\metre}$ (\texttt{R2B10} grid), and is available with an hourly output frequency.
The simulations were performed with the Icosahedral Non-hydrostatic modelling framework (ICON) model \cite{RN167,RN4}, and for each day of the $2$-month data set the simulations where initialized at $0000\,\unit{\UTZ}$ and run for $\qty{36}{\hour}$.
For this simulation the ICON model was used in its numerical weather prediction (NWP) setup without parameterizations for convection and subgrid-scale orography.
Parameterizations for radiation, cloud microphysics, and turbulence were active \cite{RN57}.
The ICON model solves the fully compressible Navier-Stokes equations with the density $\rho$ as a prognostic variable.
ICON uses an icosahedral-triangular C grid and has a non-hydrostatic dynamical core \cite{RN167}.

\begin{figure}[tbh]
    \centering
    \includegraphics{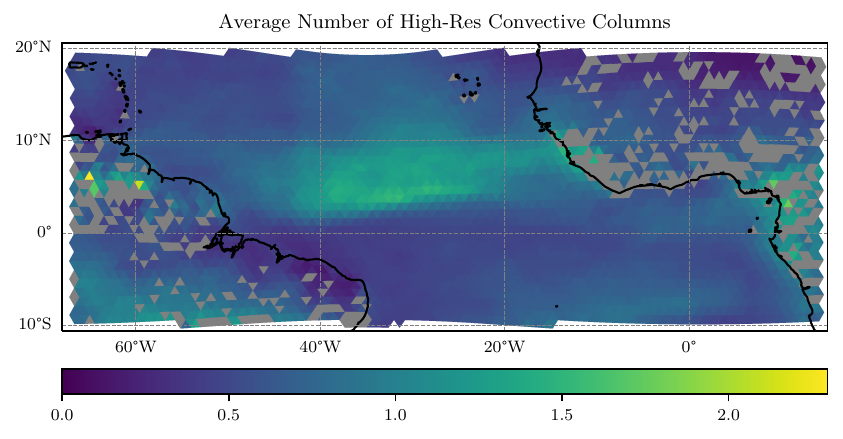}
    \caption{Average number of high-resolution convective cells per displayed low-resolution column and time frame as defined in \cref{eq:ConvectiveConditions} in the studied tropical Atlantic region over the entire considered period of time. In the west the coastline of South America and some Caribbean islands can be seen and in the east the coastline of Africa. The low resolution grid has an approximate horizontal resolution of $\Delta x\approx \qty{80}{\kilo\metre}$. Excluded columns are marked in grey.}
    \label{fig:AverageConvectiveCellsAcrossNarval}
\end{figure}

These simulations are well suited for learning a coarse-resolution data-driven convection scheme, as a high number of convective cases are present in the tropical Atlantic region. In \Cref{fig:AverageConvectiveCellsAcrossNarval} the spatial distribution of the average number of convective cells per column (as defined below) in the studied region is shown.
Columns excluded from the training dataset as described later in the coarse-graining section (\cref{sec:Coarse-Graining}) are marked in grey.
The figure shows a clear pattern of the ITCZ (compare \citeA[Fig.~2]{RN58}) with an increased number of convective cells.
Additionally, many convective cells can be found along the coast and over mountainous terrain.
While many columns over mountainous terrain are filtered out from the data set, there are still many datapoints to learn from over these areas, as seen in \Cref{fig:AverageConvectiveCellsAcrossNarval}.

As a first preprocessing step we discarded the first hour of every day in the dataset because of some discontinuous behavior at the start of each day related to the initialization/\allowbreak{}spin-up phase of the simulations.
Additionally, we also cropped the original NARVAL region by \qty{2}{\degree} on all sides since we noticed some boundary effects in the spatial patterns as well.
The region seen in \Cref{fig:AverageConvectiveCellsAcrossNarval} was already cropped by the mentioned \qty{2}{\degree}.

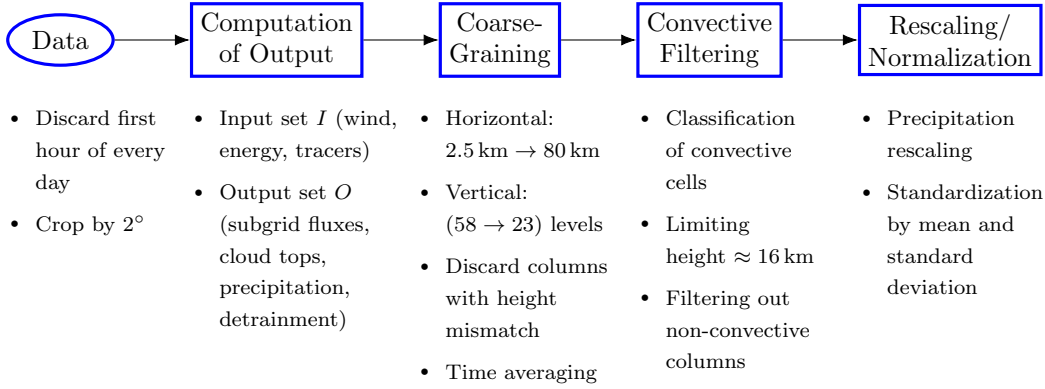
\begin{figure}
    \centering
    \begin{tikzpicture}[
    roundnode/.style={circle, draw=blue!100, very thick, minimum size=7mm, align=center},
    ellipsenode/.style={ellipse, draw=blue!100, very thick, minimum size=5mm, align=center},
    squarednode/.style={rectangle, draw=blue!100, very thick, minimum size=5mm, align=center},
    textnode/.style={align=left, inner sep=-6mm, font=\footnotesize, anchor=north west},
    ]
    %Nodes
    \node[ellipsenode] (sq1)                {Data};
    \node[squarednode] (sq2) [right=of sq1] {Computation\\of Output};
    \node[squarednode] (sq3) [right=of sq2] {Coarse-\\Graining};
    \node[squarednode] (sq4) [right=of sq3] {Convective\\Filtering};
    \node[squarednode] (sq5) [right=of sq4] {Rescaling/\\Normalization};
    % \node[squarednode] (t1)  [below=of sq1] {test}

    % Text below nodes
    \node[] (placeholder) at (sq2.west |- 0,-4.6) {}; % This node only creates some space below the figure
    \node[textnode, text width=3cm] (t1) at (sq1.west |- 0,-1.2) {
        \begin{itemize}
         \item Discard first hour of every day
         \item Crop by \qty{2}{\degree}
        \end{itemize}
    };
    \node[textnode, text width=3.5cm] (t2) at (sq2.west |- 0,-1.2) {
        \begin{itemize}
         \item Input set $I$ (wind, \allowbreak{}energy, \allowbreak{}tracers)
         \item Output set $O$ (subgrid fluxes, \allowbreak{}cloud tops, \allowbreak{}precipitation, \allowbreak{}detrainment)
        \end{itemize}
    };
    \node[textnode, text width=3.5cm] (t3) at ($(sq3.west |- 0,-1.2)+(-0.3,0)$) {
        \begin{itemize}
         \item Horizontal: $\qty{2.5}{\kilo\metre} \rightarrow \qty{80}{\kilo\metre}$
         \item Vertical: $(58\rightarrow 23)$ levels
         \item Discard columns\\with height\\mismatch 
         \item Time averaging
        \end{itemize}
    };
    \node[textnode, text width=3cm] (t4) at (sq4.west |- 0,-1.2) {
        \begin{itemize}
         \item Classification of convective cells
         \item Limiting height $\approx\qty{16}{\kilo\metre}$
         \item Filtering out non-convective columns
        \end{itemize}
    };
    \node[textnode, text width=3.1cm] (t5) at (sq5.west |- 0,-1.2) {
        \begin{itemize}
         \item Precipitation rescaling
         \item Standardization by mean and standard deviation
        \end{itemize}
    };
    
    %Lines
    % \draw[->] (sq1.east) -- (sq2.west);
    \draw[-{Latex[length=2mm]}] (sq1.east) -- (sq2.west);
    \draw[-{Latex[length=2mm]}] (sq2.east) -- (sq3.west);
    \draw[-{Latex[length=2mm]}] (sq3.east) -- (sq4.west);
    \draw[-{Latex[length=2mm]}] (sq4.east) -- (sq5.west);
    \end{tikzpicture}
    \caption{Summary of preprocessing steps. Starting from the original data, first the subgrid fluxes as well as 2D outputs, such as precipitation, were computed. After this, the data was coarse-grained and filtered for active convection. As a final preprocessing step, the data was rescaled and normalized.}
    \label{fig:preprocessing_steps}
\end{figure}

To give a short overview of the preprocessing steps described below, \Cref{fig:preprocessing_steps} depicts an overview of the various steps used, beginning with the original data set.

\subsection{Computation of Output}\label{sec:Computation of Output}

The selection of input and output variables for the ML models are based on the implementation of the cumulus scheme in the ECHAM6 model \cite{RN44,RN42,RN9}.
They correspond to the physical quantities transported by convective processes and a few related quantities such as precipitation.
If not stated differently, we used the following set of variables for the input of the convective scheme

$$I=\{u,v,w,h,q_v,q_l,q_r,q_i,q_s\}.$$

This set consists of the zonal, meridional, and vertical wind components ($u,v,w$), as well as the liquid/ice water static energy ($h$) and five different tracer species.
These tracer species are the specific humidity ($q_v$) and specific cloud water, cloud ice, rain, and snow content ($q_l,\allowbreak{}q_i,\allowbreak{}q_r,\allowbreak{}q_s$).
The liquid/ice water static energy is defined here as

\begin{equation}
    h = c_p T + z g - L_v \cdot (q_c + q_r) - L_s \cdot (q_i + q_s + q_g),
\end{equation}

with temperature $T$, altitude $z$, the specific heat at constant pressure $c_p$, specific graupel content $q_g$, and the latent heat of evaporation and sublimation $L_v$ and $L_s$.
We chose to not give the ML models any information about their spatial location or solar insolation in order to force them to learn from the dynamical state.
This also enables the application of the trained models outside of their limited training domain.

Correspondingly, the output fields are

$$O=\{F^\mathrm{sg}_u,\allowbreak{}F^\mathrm{sg}_v,\allowbreak{}F^\mathrm{sg}_h,\allowbreak{}F^\mathrm{sg}_{q_v},\allowbreak{}F^\mathrm{sg}_{q_l},\allowbreak{}F^\mathrm{sg}_{q_r},\allowbreak{}F^\mathrm{sg}_{q_i},\allowbreak{}F^\mathrm{sg}_{q_s},z_{cltop},\allowbreak{}p_{cltop}, \allowbreak{}q_{l,detr},\allowbreak{}q_{i,\mathrm{detr}},\allowbreak{}P\}.$$

The first eight variables with notation ``$F^\mathrm{sg}_{var}$'' are 3D fields and correspond to the subgrid flux component of the input variables $I$ (excluding $w$).
The remaining variables in the output set are 2D fields, namely cloud top height ($z_{cltop}$), cloud top pressure ($p_{cltop}$), integrated liquid/ice detrainment ($q_{l,\mathrm{detr}}$, $q_{i,\mathrm{detr}}$), and precipitation ($P$).
For the cloud top level we chose to predict the altitude as well as the pressure, although they contain very similar information, because our goal was to provide the same output as the ECHAM6 cumulus scheme.

We focused on predicting subgrid fluxes instead of the direct tendencies because this allowed  abiding conservation laws by applying appropriate boundary conditions (no-flux at the top and a flux which is consistent with the surface forcing at bottom).
We decomposed variables such as the density ($\rho$) into a horizontal spatial average (on the same model level) over the coarse resolution, denoted by an overline, and a fluctuating component, denoted by a prime, as $\rho=\overline{\rho}+\rho^\prime$.
The fluctuating component therefore represents the departure from the coarse grid average.
This enabled us to calculate the subgrid (i.e., unresolved) vertical advective flux of, say, the variable $u$, $F^\mathrm{sg}_u$, for a given coarse resolution as follows:

\begin{equation}
    \label{eq:sgf_computation}
    F^\mathrm{sg}_u = \overline{\rho w u} - \overline{\rho}\,\overline{w}\,\overline{u} = \overline{\rho}\,\overline{w^\prime u^\prime}+\overline{w}\,\overline{\rho^\prime u^\prime}+\overline{u}\,\overline{\rho^\prime w^\prime}+\overline{\rho^\prime w^\prime u^\prime}.
\end{equation}

This subgrid momentum flux $F^\mathrm{sg}_u$ was calculated as the difference between the coarse-grained flux $\overline{\rho w u}$ obtained by first calculating the flux with the high-resolution resolved variables, then coarse-graining it to the coarser resolution, and the flux calculated with the low-resolution variables $\overline{\rho}\,\overline{w}\,\overline{u}$ (see \cref{eq:sgf_computation}).
The term on the right hand side in \cref{eq:sgf_computation} results from the fact that averages over fluctuations are by definition zero.
This method is similar to the one of \citeA{RN25}, but without neglecting the horizontal density fluctuations between high-resolution cells within a coarse resolution target cell of the coarse-graining procedure.
This is especially important for models with terrain-following vertical coordinates, such as the height based terrain following vertical coordinate of the ICON model \cite{RN4}, because horizontally neighbouring cells (same vertical level) in the lower troposphere over land with steep topography can have strongly different height, thus different pressure and density.
By looking into the subgrid variations of $\rho$ we found that, especially in the lowest levels over heterogeneous terrain, there are fluctuations of up to \qty{25}{\percent} of the mean value within a single coarse grid cell.
As we are calculating the subgrid flux from a single snapshot of the dynamics and do not consider differences between timesteps, the subgrid flux represents the flux difference between the coarsened high-resolution state and coarse state due to resolved processes.
Here, these resolved processes are cumulus convection and gravity waves since we only learn from convective columns (method is shown later in this section).
Gravity wave drag mainly impacts higher levels \cite{Kim2003-overview-gwd} and the here developed parameterizations are limited in height (see \Cref{fig:ConvectiveCellProfile}).
The momentum flux due to gravity waves, excited by convection, is a second order effect which we neglect here.%so that the resulting subgrid flux is assumed to be due to convection.

For the cloud top height/cloud top pressure ($z_{cltop}$/$p_{cltop}$) we took the height/\allowbreak{}pressure of the highest cell with convective clouds found according to the condition formulated in the next section (\cref{eq:ConvectiveConditions}).
While there are different ways to estimate the detrainment of liquid/ice \cite{RN59,zhang2019sensitivity} we decided to follow \citeA{RN42,RN50} and calculated the fractional detrainment as

\begin{equation}
    \delta = -\frac{1}{\sigma}\frac{\partial \sigma}{\partial z},
\end{equation}

where $z$ is the altitude and $\sigma$ the fractional cloud area.
As such, it was possible to calculate the integrated detrainment of water and ice by multiplication with the vertical mass flux and integrating along the column.
Before integration, the column was masked according to its temperature (above or below \qty{0}{\celsius}) \cite{RN9} to differentiate between liquid and ice detrainment.
For precipitation we cannot assume that it stems entirely from convective precipitation in convective columns as stratiform and convective precipitation often occur simultaneously \cite{houze_strat-conv-prec-paradox,schumacher-funk_assesing-conv-strat-prec}.
Therefore, when coupling the ML parameterization to the ICON model we will set the large-scale precipitation from the model to zero in regions where the ML parameterization is active.
Another approach would be to classify the precipitation in the high-resolution data as convective or not, based on thresholds on e.g. vertical velocity, precipitation rate or based on the spatial structure of precipitation clusters.
Here, we decided to predict both precipitation types together as the before mentioned approaches would introduce additional degrees of freedom into the method and therefore complexity.

\subsection{Coarse-Graining}\label{sec:Coarse-Graining}

The coarse-graining was done first in the horizontal and afterwards in the vertical direction as described in \citeA{RN24} for a data-driven cloud cover parameterization.
The horizontal coarse-graining from the \texttt{R2B10} ($\Delta x\approx\qty{2.5}{\kilo\metre}$) to an \texttt{R2B5} ($\Delta x\approx\qty{80}{\kilo\metre}$) grid was performed with the help of the \texttt{remapcon} function from the Climate Data Operators (CDO) \cite{schulzweida_uwe_2022_7112925}.
At this scale individual convective clouds and smaller convective systems are coarse-grained, allowing us to parameterize their average impact on the large-scale dynamics.
In the vertical, we reduced the resolution from $58$ to $23$ levels up to the mentioned limiting height of $\sim\qty{15.9}{\kilo\metre}$ in \Cref{fig:ConvectiveCellProfile}.
The vertical coarse-graining operator works in a similar way as the horizontal averaging.
The high-resolution cells were averaged weighted by their fractional proportion in the coarse cell \cite{RN24}.
Some low-resolution columns have a significantly lower base than the high-resolution cells because of the more detailed topography in the high-resolution data.
Therefore, it was not possible to compute reasonable averages with the above described coarse-graining operator in the lowest model levels.
Here, we also adopted the method from \citeA{RN24} and excluded columns with a significant difference between the vertical extent of low and high-resolution columns of the dataset.

In the high-resolution data, cells on the same vertical level can be on different geometric heights due to the terrain-following coordinate system. An approximation applied here is that the coarse-graining is first performed in the horizontal and afterwards in the vertical. Therefore the result can be different from coarse-graining over the low-resolution volume \cite{RN24}.

Additionally, we introduce time-averaging to reduce the noise from instantaneous snapshots of the dynamics as it was found to reduce model overfitting in \citeA{2020arXiv201012559R}.
For a column in the data set at time $t_i$, we average the column variables and fluxes over the time steps $t_{i-1},\allowbreak{}t_{i},\allowbreak{}t_{i+1}$, corresponding to a moving window of a three-hour duration.
Physically, the three-hour temporal averaging should still allow to resolve the life cycle of the tropical deep convective clouds with a diurnal cycle \cite{chen1997-diurnal_variation_and}.
A \qty{3}{\hour} window is just about short enough to resolve the life cycle of such clouds and still allow a minimal smoothing of higher frequency variability.

\subsection{Filtering for Convection}\label{sec:Filtering for Convection}
In order to learn mainly from columns in which convection has a dominant impact on the overall dynamics we introduced a filtering of the data.
First, individual high-resolution cells were classified as convective if the following conditions \cite{RN67,RN166} are met:

\begin{equation}
    \label{eq:ConvectiveConditions}
    q_l+q_i > \qty{0.01}{\gram\per\kilo\gram},\quad w > 0,\quad B\propto \theta_v - \overline{\theta_v} > 0,
\end{equation}

where $w$ is the vertical velocity, $\theta_v$ is the virtual potential temperature, and $q_l$/$q_i$ are the specific cloud liquid water/cloud ice content, respectively.
Additionally, the buoyancy $B$ has been introduced in the conditions (\ref{eq:ConvectiveConditions}).
In this case the overline denotes horizontal averaging over approximately \qty{10}{\kilo\metre}.
We chose this averaging scale as convection becomes partly resolved by grid scale dynamics for resolutions higher than approximately \qty{10}{\kilo\metre} \cite{Ahn2018,acp-11-3731-2011}.
The averaging was performed with the \texttt{remapcon} function \cite{schulzweida_uwe_2022_7112925} to an R2B8 resolution.
Next, we classified entire low resolution columns as convective or non-convective.
For this, the number of convective cells per high-resolution column was summed up along the height dimension and coarse-grained horizontally (as explained above).
If the so-calculated 2D field was equal (or higher than) 1 for a given column, so that on average all high-resolution columns inside the coarse column had at least one convectively classified cell, this coarse column was classified as convective and was added to the training data set.
These columns are henceforth referred to as ``convective'' columns.
A time average over the entire observed period of this so computed low resolution data is displayed in \Cref{fig:AverageConvectiveCellsAcrossNarval}.
Furthermore, we added $\qty{10}{\percent}$ of the non-convective columns for training so that we ended up with slightly more than \qty{2}{million} coarse sample columns.
Before the filtering, there were about \qty{5}{million} low-resolution and approximately \qty{455}{million} high-resolution columns in the whole data set.

\begin{figure}[tbh]
    \centering
    \includegraphics{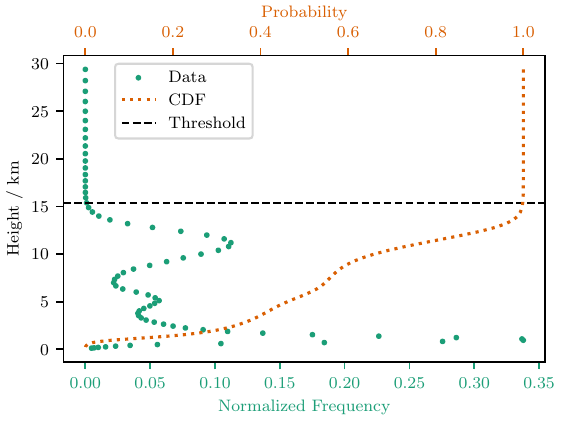}
    \caption{Probability distribution of convectively classified cells over altitude (green large dots) in the high-resolution data with ICON over the NARVAL region. The orange dashed line shows the cumulative distribution function (CDF) and the black dashed line represents the height up to which \qty{99.9}{\percent} of the convective cells are found. The bottom scale corresponds to the probability and the top to the cumulative distribution.}
    \label{fig:ConvectiveCellProfile}
\end{figure}

In order to find a limit in altitude to predict unresolved convective effects, we considered that convection in the atmosphere under normal conditions is limited by the tropopause \cite{RN161}.
Therefore, we checked up to which height we find convectively classified cells in the data set.
The result can be seen in \Cref{fig:ConvectiveCellProfile}.
The limiting height in the figure is drawn at the height up to which \qty{99.9}{\percent} of the convectively classified cells are found (compare dashed orange line).
This height is at $\sim\qty{15.9}{\kilo\metre}$, which is reasonable considering the tropical tropopause height of roughly $\qtyrange{12}{17}{\kilo\metre}$ \cite{RN146}.
Only values below this height are considered as input and output to the machine learning algorithms.

The general form of the data observed in \Cref{fig:ConvectiveCellProfile} resembles the expected trimodal distribution of convective clouds in the tropics \cite{RN157}.
The lowest peak corresponds to (shallow) cumulus, the peak at $\sim\qty{5}{\kilo\metre}$ to cumulus congestus and the highest clouds found are deep cumulonimbus clouds.

\subsection{Rescaling and Normalization}\label{sec:Rescaling and Normalization}

For higher numerical stability of the machine learning models and to have the variables on the same scale, we standardize the 2D fields by subtracting the mean across samples from all 2D variables and dividing by the standard deviation.
The same procedure is done for all 3D variables, but in this case mean and standard deviation are calculated across the height dimension as well.
We also tested normalizing variables by their mean and standard deviations level by level but observed a decrease in model skill.

Furthermore, before applying the standardization, we use the following nonlinear rescaling for the accumulated precipitation $P$ per hour:

\begin{equation}
    P^\prime = \ln{\left(1+\frac{P}{\qty{1}{\kilo\gram\per\metre\squared\per\hour}}\right)}.
\end{equation}

The reason for this is that precipitation intensities are typically represented by a heavily skewed (gamma) distribution \cite{RN162}.
This distribution is characterized by a comparatively large number of low values and very few heavy precipitation events.
Without a proper rescaling, ML models would achieve a low prediction error by predicting zero precipitation regardless of the input \cite{RN60}.
Additionally, it is well known that coarse GCMs have a bias towards low intensity precipitation events \cite{RN151,RN19}.
The rescaling should help mitigate some of this problem.

\section{Machine Learning Models} \label{sec:ml_methods}
As mentioned in the introduction, ML-based convection parameterizations have been developed using different kinds of methods.
These include RFs \cite{RN2,RN17,RN176}, MLPs \cite{RN5,RN19,RN25,iglesias-suarez_causal_nns}, ensembles of MLPs \cite{RN127}, Residual Convolutional Neural Networks \cite{RN163,han-an-ensemble-NN-moist-physics}, Residual Neural Networks (ResNets) \cite{RN125}, Generative Adversarial Networks (GANs) \cite{RN130}, and Variational Encoders (VAEs) / Variational Auto Encoder Decoders (VEDs) \cite{RN140,RN88}.
One goal of this study is to evaluate various kinds of machine learning models on the same data set.
Therefore, we first introduce the used models.
All models use a vertical column (23 height levels and nine variables) from the sample data set as input and the column fluxes (23 height levels and eight variables) plus five 2D variables as output, see above.

We tested four different deep learning architectures: Multilayer Perceptron, Convolutional Neural Network (CNN), Residual Neural Network \cite{RN173}, and a convolutional neural network with a U-shaped architecture (U-Net) \cite{RN91}.
The MLP family consists of several fully connected layers with additional optional batch normalization layers and activation functions (see section \ref{sec:appendix_hpo}).
Furthermore, we introduced a linear model (LinMLP) which is based on the best found architecture of the MLP class but all nonlinear activation functions are replaced by linear ones.
For the CNN class we decided to consider networks with a first convolutional layer connected to some number of fully connected layers thereafter.
All convolutions are 1D convolutions in the vertical as the data set consists of variables on different levels due to the typical neglect of horizontal interactions and variability for parameterized processes in climate models.
The ResNet architecture is inspired by \citeA{RN125}, the network consists of several different blocks with some number of fully connected layers and optional batch normalization.
The input of each block is added to its output to form the final output set.
This helps prevent vanishing gradients and degradation \cite{RN173}.
For the gradient-based optimization of the networks we chose to use the Adam algorithm \cite{2014arXiv1412.6980K}.
For the implementation of all deep learning models we relied on the Pytorch library \cite{Paszke_PyTorch_An_Imperative_2019}.

\begin{figure}%[tbh]
    \centering
    \includegraphics{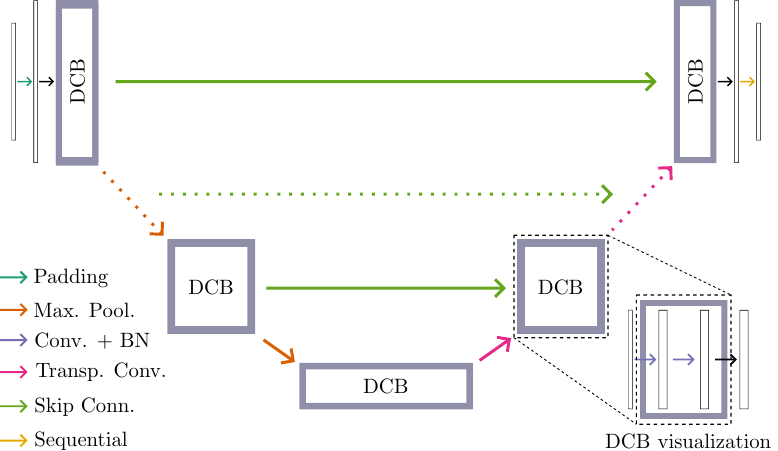}
    \caption{Visualization of the used U-Net architecture. The abbreviations DCB, Conv., Transp. Conv., and BN stand for double convolution block, convolutional layer, transpose convolutional layer, and batch normalization layer, respectively. The dotted lines mark the possibility for more blocks depending on the result of the hyperparameter optimization (HPO). The horizontal lines indicate skip connections. In the lower right of the figure, a more detailed visualization of the double convolution block is given.}
    \label{fig:UnetVisualization}
\end{figure}

Furthermore, we decided to use a U-Net architecture, see \Cref{fig:UnetVisualization}.
This network is similar to the ResNet in the sense that it contains residual connections and that it is constructed out of structurally similar blocks.
In contrast to the ResNet, these blocks use two convolutional layers each instead of an arbitrary number of fully connected layers.
Additionally, this architecture utilizes max pooling and transpose convolution layers to compress and expand the input in the height dimension.
This allows the network to process the input information on multiple spatial scales.
During the compression process (left part of \Cref{fig:UnetVisualization}) the channel dimension (width in the figure) grows.
The kernel size of the convolutions stays constant but the height dimension shrinks, this effectively increases the receptive field for each consecutive layer in the network.
The U-Net is therefore able to detect patterns on scales between the models vertical level spacing (\qty{\sim 30}{\metre} at the lowest level or up to \qty{\sim 500}{\metre} for the highest predicted level) and the column height (\qty{\sim 16}{\kilo\metre}).
In the expansion process (right part) the channel dimension shrinks again.
We propose this architecture, which is particularly suited for multiscale modeling, for the given parameterization problem because of the multiscale nature of moist convection \cite{RN147}.
The U-Net has favorable properties for our problem as local features can be picked up by the network on a variety of different scales throughout the downscaling process, and the residual connections help to communicate this information to the upscale branch of the network.
This capability is crucial for tasks that require understanding both local and global context within the input data, such as in image segmentation \cite{RN91} where the target output can depend on patterns of varying sizes and resolutions.
In the context of convection, the initial layers are capable of capturing more small-scale convective systems/flows and the more compressed layers are responsible for representing deep convection/large-scale systems.

Besides these deep learning architectures, we trained five different non-deep learning models.
For the implementation of these we used Scikit-Learn \cite{scikit-learn}.
As lowest complexity models we used linear methods such as Lasso \cite{10.1111/j.2517-6161.1996.tb02080.x} and Ridge \cite{doi:10.1080/00401706.1970.10488634} regression.
Additionally, we used three tree-based models.
These include Random Forests (RF) \cite{Breiman2001}, Extra Trees (ET) \cite{Geurts2006}, and Gradient Boosted Trees (GBT) \cite{FRIEDMAN2002367}.
Further information about the different ML models can be found in section \ref{sec:appendix_nondl_info}.

To select an appropriate set of hyperparameters we chose to split the data non-consecutively into a training/validation/test set with a fraction of \qty{80}{\percent}/\qty{10}{\percent}/\qty{10}{\percent} of the data.
This corresponds to $\sim 1.6\,\cdot\,10^{6}$ sample columns for training and $\sim 2\,\cdot\,10^5$ columns for validation/testing.
Depending on the architecture we treated the different input variables as separate channels (for CNN and U-Net) and otherwise concatenated them in one vector.
The output variables were always concatenated in one vector.
For the non-deep learning algorithms we first did the hyperparameter optimization (HPO) on a subset of the data from five random days ($\sim 1.6\cdot 10^5$ samples) because most of the models have difficulties with handling vast amount of data.
The models identified as best in the HPO where then trained on the whole data set.
An explanation of the different hyperparameters involved in all models can be found in section \ref{sec:appendix_hpo}.

\section{Results}\label{sec:results}% (Or section title might be a descriptive heading about the results)
This section will first introduce a model evaluation for all ML models used and then focus on a more detailed comparison of the highest performing (offline) deep and non-deep learning method in \cref{sec:model_evaluation}.
Afterwards, in \cref{sec:explainability_of_u-net_and_gbt}, we will investigate what the models have learned and find that, in fact, an ablated version of the U-Net (without precipitating tracers as input) learns physically explainable relations as opposed to the non-ablated version.
This ablated model, in comparison with the non-ablated version, will also be tested in the online stability test section in the end of this chapter in \cref{sec:online_stability_tests}.

The architecture of the best performing model, the U-Net, is first introduced in \cref{sec:model_evaluation} and the ablation, improving online stability, is described in \cref{sec:explainability_of_u-net_and_gbt}.

\subsection{Machine Learning Model Benchmarking}\label{sec:model_evaluation}

First, we focus on the simple aggregated evaluation of the coefficient of determination ($R^2$) values for all examined model classes.
The $R^2$ value is calculated as 1 minus the mean squared error of the predictions over the variance of the data.
We compute the $R^2$ value across variables and levels, a more detailed (per variable/level) comparison is given later in \Cref{fig:R2HeightProfiles}.
All models have been hyperparameter-tuned according to the method described below.
Briefly this HPO consisted of running a large ensemble of models with parameters sampled from predefined search spaces and their performance evaluated on a validation set (more details in section \ref{sec:appendix_hpo}).

\begin{figure}[tbh]
    \centering
    \includegraphics{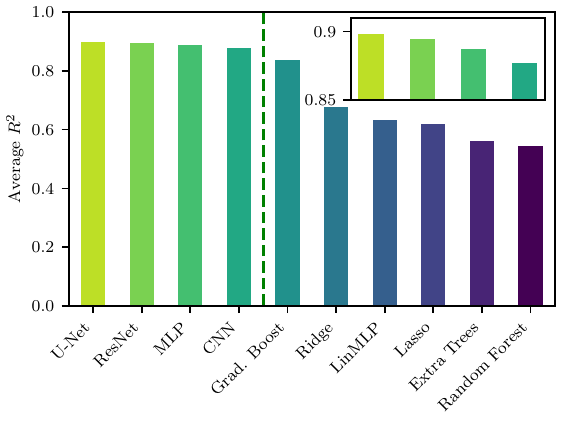}
    % \scalebox{0.48}{\input{ModelR2Comparison.pgf}}
    % \resizebox{1\linewidth}{!}{\input{ModelR2Comparison.pgf}}
    \caption{Coefficient of determination ($R^2$) on a test set for different types of models. All models were hyperparameter-optimized, and the best models were then trained on the whole data set. The deep learning methods are displayed on the left of the green dashed line and the non-deep learning methods on the right of it. The inset in the top right shows a zoomed-in version of the $R^2$ for the deep learning models.}
    \label{fig:ModelR2Comparison}
\end{figure}

\Cref{fig:ModelR2Comparison} displays the $R^2$ values for all models over all variables and levels.
On the left hand side of the dashed green line the deep learning models are shown as opposed to the simpler models on the right hand side.

The $R^2$ value of the Random Forest is the lowest of the examined models.
RFs have been used as data-driven convection parameterizations with some success \cite{RN2,RN17} in idealized settings before.
Limitations in the application of RFs for realistic parameterization schemes have been observed before due to their computational inefficiency, memory requirements, and comparably low complexity (versus deep neural networks for instance), limiting their capacity to capture high dimensional features \cite{RN176}.
The GBT model class has a strikingly high $R^2$ value, comparable to the ones of the deep learning methods.
This suggests that these RF-based parameterization schemes could improve in performance if they were based on Gradient Boosted Trees (besides deep learning networks).
The Extra Trees model has a similarly low performance as the RF.
Considering that the ET model is structurally similar to RFs, including an additional element of randomness as explained above, this is not surprising.
The linear models (Ridge, LinMLP, and Lasso) show relatively high performance compared to that of the RF/ET model with $R^2$ values of $0.68,0.63,0.62$.
The $L^2$-regularization term seems to have a higher impact on the generalization capabilities of the linear model compared to the $L^1$-regularization in Lasso regression.
The generally better performance of the linear models compared to the tree based models, RF and ET, is surprising and might be connected to the fact that linear models are able to extrapolate to unseen data points based on the linear relationships learned during training.
These tree based methods, however, are limited to the range of the training data and cannot extrapolate beyond it because they predict based on averages of similar seen samples.
As we are using high-dimensional data some degree of extrapolation is very probable \cite{balestriero2021-highdim-extra}.
Another point is that in cases of high-dimensional data with many uninformative or noisy features, linear models, especially when combined with regularization techniques like Lasso, can perform better by effectively reducing the dimensionality and focusing on the most relevant features.
Random Forests might not be as effective in ignoring these irrelevant features to that extent.
Another option might be that the linear models are being too heavily tuned to the tropical convection problem.
More on this in the discussion.

\begin{figure}%[tbh]
    \centering
    \includegraphics{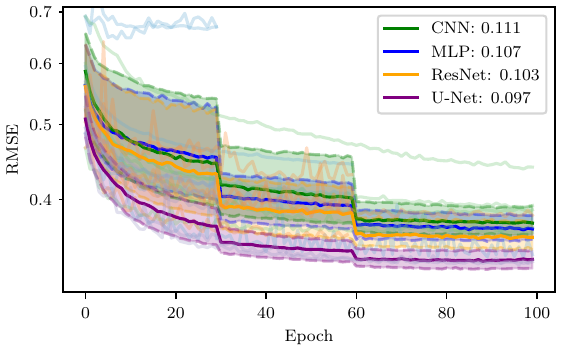}
    \caption{Root mean squared error during HPO on the validation set of the four different deep learning methods. The straight thick lines correspond to the median of the HPO ensemble, the shaded areas are drawn in between the first and third quartile. Additionally, ten realizations for each DL method are shown in similar colors. The legend shows the minimum of the validation loss for each of the methods. The scheduler of the HPO filters badly performing runs after 30 and 60 epochs, causing the steps in the profiles. For this task we used the \texttt{AsyncHyperBandScheduler} \cite{2018arXiv181005934L} of the \texttt{Ray Tune} library \cite{liaw2018tune}.}
    \label{fig:NNTestLossCurves}
\end{figure}

The deep learning models outperform the other methods but, e.g., for the GBT model only by a small amount.
While the $R^2$ value for the GBT is almost as high as the value for the U-Net, the other nonlinear methods show a rapid decrease in performance when ordering by their respective $R^2$ value.
\Cref{fig:ModelR2Comparison} shows that the performance difference between the various deep learning models measured by $R^2$ is negligible.
One could suspect that the best performance of the U-Net could originate purely by chance.
Therefore, we performed an extensive HPO with over $5000$ ensemble members in total.
The resulting median/upper/lower quartile profiles can be seen in \Cref{fig:NNTestLossCurves}.
We varied hyperparameters such as the learning rate, number of neurons/layers/blocks, or activation functions.
More details on the HPO search spaces can be found in section \ref{sec:appendix_hpo}.
A visualization of the HPO and the training and validation process in general is shown in Figure \ref{fig:HpoVisualization}.
We notice that the U-Net has a consistently lower error than the other models, and the upper quartile of its distribution is on the same level as the lower quartile of the second best performing model, the ResNet.
The difference between the other model classes is smaller, and the spread around each median profile is larger than for the U-Net.

Furthermore, the model complexity of the U-Net is comparatively low.
As it can be seen in the number of parameters of our network configurations (Table \ref{tab:net_num_params}) and Figure \ref{fig:ComplexityPerformance}, the most complex (judging by number of parameters) deep learning model is the ResNet with more than four times the number of parameters of the U-Net.
The MLP architecture has the lowest number of parameters, the U-Net has the second lowest number before the CNN and ResNet.
Despite this, the U-Net shows the consistently lowest error on the validation/test set (see \Cref{fig:NNTestLossCurves}) over a large set of hyperparameter configurations, presumably because of its multiscale architecture and the resulting ability to capture multiscale problems such as convection well.

\begin{figure}[tbh]
    \centering
    \includegraphics{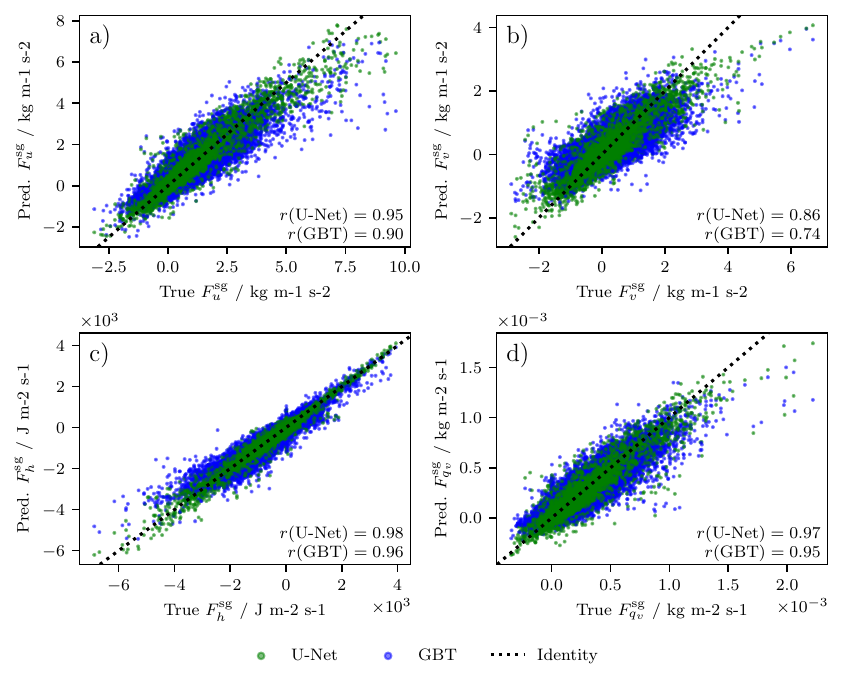}
    \caption{Scatter plot for the subgrid fluxes of a) zonal, b) meridional momentum, c) liquid/ice water static energy, and d) specific humidity. Data for the U-Net is shown in green, for the GBT in blue, and the diagonal is marked by a dotted line. The Pearson correlation coefficient $r$ between the true and the predicted subgrid flux is noted in the lower right corner of each plot for both U-Net and GBT.}
    \label{fig:SubgridFluxScatter_uvhqv}
\end{figure}

\begin{figure}[tbh]
    \centering
    \includegraphics{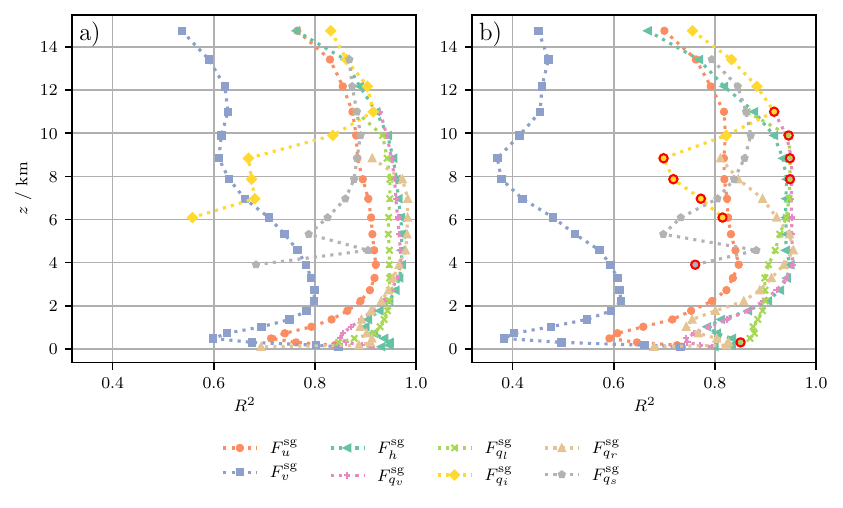}
    \caption{Average $R^2$ profile for all subgrid flux variables for a) the U-Net and b) GBT model. Data points where the GBT model actually has a higher $R^2$ than the U-Net are additionally marked by a red circle.}
    \label{fig:R2HeightProfiles}
\end{figure}

Based on these results, we will focus on the respectively best performing deep and non-deep learning models from now on.
These models are the U-Net and the GBT model as seen in \Cref{fig:ModelR2Comparison}.
We first compare the U-Net and GBT flux predictions with the true values for $F^\mathrm{sg}_u,\allowbreak{}F^\mathrm{sg}_v,\allowbreak{}F^\mathrm{sg}_h,\allowbreak{}F^\mathrm{sg}_{q_c}$ over all levels.
The results can be seen in \Cref{fig:SubgridFluxScatter_uvhqv}, and a corresponding plot showing the distribution for the remaining tracer subgrid fluxes can be seen in Figure \ref{fig:SubgridFluxScatter_qlqiqrqs}.
The correlation is always higher for the U-Net predictions, and for both models the meridional momentum fluxes are the hardest to predict.
This has been noted before e.g., for a data-driven gravity wave scheme \cite{espinosa2022-MlGravityWaveParam}.
The diurnal cycle and its annual variability are typically more pronounced \cite{rs14030459} for the meridional wind and can be be out of phase in the northern and southern hemisphere \cite{Ueyama2008-AClimatologyofDiurnal}.
We assume that, therefore, it is a challenge for the ML models to predict the meridional momentum flux receiving as input only the large-scale state, which my not adequately represent the nuances of meridional dynamics.

Especially for high values of the flux, both models tend to underestimate the true flux, which can be seen by the points below the diagonal in plot b).
To a similar extent, this trend can also be seen for the fluxes $F^\mathrm{sg}_u$ and $F^\mathrm{sg}_{q_v}$.
The mentioned fluxes of the GBT show a slight corresponding overestimation for low flux values.
In contrast to that, the U-Net data distribution is more symmetric about the main diagonal.
This means that there is no or a very small systematic under- or over-prediction for these values by the U-Net.
In general, the spread around the diagonal is bigger for the GBT than for the U-Net.
This confirms the better performance of the U-Net seen in \Cref{fig:ModelR2Comparison} based on $R^2$ values.

After having examined the model performance aggregated over all levels we now look at the average $R^2$ values of the 3D variables on individual vertical levels.
This is shown in \Cref{fig:R2HeightProfiles}, again for the U-Net and GBT.
Some vertical levels are not shown in the figure because the variation of the variables on these levels is close to zero.
We determined the variables for which this is true by first finding the $99$th percentile of their absolute values.
Then, for each variable all levels in which the computed percentile is below \qty{1}{\percent} of the maximum percentile for the variable were excluded from the plot.

This method filters all levels which show significantly less variation compared to all other levels.
Looking at \Cref{fig:R2HeightProfiles} we filtered out the lower tropospheric values for the ice and snow tracers as well as the higher tropospheric values for cloud water and rain tracers.
This is reasonable because we do not expect much snow/ice in the lower troposphere of the tropics, and similarly, the temperatures are too low for cloud water/rain to exist close to the tropopause.

Comparing the plots in \Cref{fig:R2HeightProfiles}, the two models show similar patterns as seen, e.g., for the $F^\mathrm{sg}_v$ curve, but the GBT curves are mostly shifted towards lower $R^2$ values compared to the U-Net.
For most variables we find a clear advantage of the U-Net in the upper layers and around the height of the planetary convective boundary layer at $\sim\qty{1}{\kilo\metre}$.
Other than for tracer species on levels in which the corresponding concentration is typically very low, the models show difficulties to predict the subgrid momentum fluxes compared to other variables, as is particularly visible for $F^\mathrm{sg}_v$.
For subgrid momentum transport in general this has been noticed before in \citeA{RN145}.
This problem could arise from the fact that the sign of the subgrid convective momentum flux depends on the nature of convective organization \cite{RN145,RN170}, which is not resolved in the coarse data.
A few points are marked by red circles, which correspond to higher $R^2$ value for the GBT.
Most of these are close to the $R^2$ U-Net value (within an $R^2$ relative deviation of \qty{1.5}{\percent}) except for the low ice and snow tracer values.
Here we assume that the GBT shows an increased performance due to the small number of training data and its lower model complexity.
Using the U-Net increases the mean $R^2$ value of all variables.
The highest improvement by using the U-Net instead of the GBT can be seen for $F^\mathrm{sg}_v$ with an average $R^2$ improvement of $0.19$ and the second highest for $F^\mathrm{sg}_u$ with a gain of $0.09$.
In the vertical, the highest average increase in skill is observed in the boundary layer.
On these lower model levels, the dynamics are typically more complex/turbulent and therefore the higher model complexity of the U-Net is especially beneficial.
This complexity in the planetary boundary layer arises from different mechanisms such as direct surface forcings, e.g., heat and moisture flux to/from the atmosphere as well as surface drag.
Also, the dynamics are inherently more turbulent because of large wind velocity gradients and shear.
Furthermore, diurnal variations and therefore general variability are much higher close to the surface layer than in the upper troposphere/\allowbreak{}atmosphere due to the direct surface interaction.

The 2D fields are also predicted more skillfully by the U-Net, the $R^2$ values for all five predicted 2D variables are higher for the U-Net than for the GBT.
As an example, the true and predicted precipitation distribution is shown in Figure \ref{fig:PrecipitationDistribution}.
Even though the $R^2$ values for precipitation are similar ($0.897$ vs. $0.860$), the U-Net predicts the extremes of the distribution much more accurately.
For instance, the 95th percentile of the true distribution and the predicted distributions of U-Net and GBT are approximately \qty{22.28}{\milli\metre\per\hour}, \qty{19.75}{\milli\metre\per\hour}, and \qty{16.83}{\milli\metre\per\hour}.
This shows that the U-Net captures the high precipitation cases much better than the GBT.

Looking at the spatial distribution of the normalized RMSE across all variables (see Figure \ref{fig:NormRmseSpatialComparison}) we notice that both models have a lower error in the region of the ITCZ and an increase in error towards higher latitudes.
This reflects the difference in the abundance of training data as seen in \Cref{fig:AverageConvectiveCellsAcrossNarval}.

\subsection{Explainability of U-Net and GBT}\label{sec:explainability_of_u-net_and_gbt}

Having looked into the prediction results we now want to find out what the models actually have learned in order to predict the parameterization output.
This will be based on the SHapley Additive exPlanations (SHAP) \cite{NIPS2017_7062} library which analyzes ML model predictions using a game theoretic approach.
A SHAP value $shap(x=x_0, y)$ gives the deviation in an output variable $y$ due to a specific value $x_0$ of the variable $x$ from the average prediction of $y$ over a given data (sub)set $\mathcal{X}_b$.
We used the \texttt{Deep\-Explainer} class \cite{NIPS2017_7062} as an efficient explainer for deep neural networks, and the \texttt{Tree\-Explainer}/\texttt{Kernel\-Explainer} class for decision tree-based models such as GBT.

\begin{figure}[tbh]
    \centering
    \includegraphics{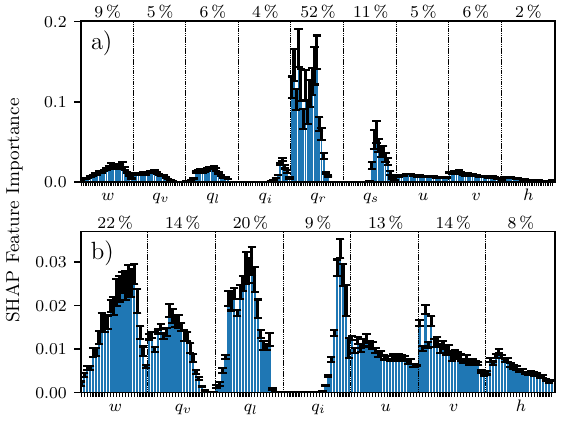}
    \caption{Feature importances (i.e., the mean absolute values of the calculated SHAP values) of input variables for a) the full U-Net model, and b) the ablated (without $qr$,$qs$) U-Net model. The mean feature importance is visualized by the height of the bar, and the standard deviation over five different computations by the errorbars. The x-axis shows different height levels for each variable, increasing from left to right. Vertical lines separate the variables. The integrated fraction of feature importances over all vertical levels is written above each variable range.}
    \label{fig:ShapImportanceComparison}
\end{figure}

\Cref{fig:ShapImportanceComparison} a) shows the mean absolute values of the calculated SHAP values for the U-Net model.
These correspond to feature importances and in this case show that the model mainly focuses on using the precipitating tracer species to predict the subgrid fluxes.
The top plot shows that $q_r$ dominates the importance attribution with over \qty{50}{\percent} of all values.
As second most influential feature we see $q_s$, another precipitating tracer species, even though it is only highly influential in the upper layers.
Additionally, one notices that the standard deviation is relatively large for $q_r$/$q_s$, indicating the ambiguity of the learned relations.
This is a first hint that the model learned non-causal relationships between convective precipitation and convective subgrid fluxes.
When the model ``sees'' coarse-grained precipitation in the data it predicts that convective subgrid fluxes must be present.
This behavior can also be observed in a more detailed analysis of the SHAP values (Figure \ref{fig:EnsembleMeanShapRhoFluct}).
Learning this connection is consistent as the link between convective precipitation and convective fluxes in the tropics is especially pronounced.
Nevertheless, this represents a weakness and non-causal link as the ML parameterization would never/rarely encounter convective precipitation in a coupled setting if it would not predict the effect of convective fluxes before.

To prevent the model from learning these non-causal connections we trained another set of models with less input variables.
We left out the precipitation input tracer species $q_r$ and $q_s$.
For this ablated model versions we performed a new HPO. These models will be discussed henceforth.
The $R^2$ performance of both models (U-Net and GBT) on the test set decreases marginally, by $\sim 0.03$, by ablating the precipitating tracers as inputs.
A third HPO was performed neglecting horizontal density fluctuations, with the result that the validation error increased for all model classes by about \qty{4}{\percent}, and for the MLP only negligibly.
This is a hint that the irreducible error of the models increases by neglecting density fluctuations.

The feature importances for the ablated U-Net are displayed in plot b) of \Cref{fig:ShapImportanceComparison}.
A more spread-out feature importance assignment can be seen in this plot: the difference between highest and lowest valued feature is only \qty{14}{\percent} which is much less than \qty{50}{\percent} as before.
This model now does not rely on spurious correlations between precipitation and convective subgrid fluxes and should generalize better outside the training domain.
The general trend for most variables seen in \Cref{fig:ShapImportanceComparison} indicates that the model focuses more on the lower model levels, and the importance is decreasing with height.
For $w$, $q_l$, and $q_i$ this is not the case, the feature importance peaks at higher model levels.
The specific cloud ice content is only present at higher altitudes as already discussed.
For the cloud water content we have very low concentrations at low model levels as clouds generally form in the boundary layer during daytime \cite{stull1988introduction}, and the mean vertical velocity profile also shows higher values at greater altitudes, indicative of the importance of shear such as on mesoscale convective system organization \cite{rotunno1988}.

We looked at the feature importance in \Cref{fig:ShapImportanceComparison} but did not discuss the influence of an input on the various output variables.
For this, we now first explain the method and then discuss the results.
For ease of notation, we focus here on a single output model with output variable $y$ as before, but this can easily be generalized to higher dimensional output.
To get the average effect of an input variable $x_i$ on the output variable $y$ we first define the fluctuation of $x_i$ for sample $j$ as $x_{ij}^\prime = x_{ij} - \langle x_i\rangle$, where the brackets $\langle\cdot\rangle$ denotes the average value over $x_i$ in the set $\mathcal{X}$.
The data set $\mathcal{X}$ is a random subset of the whole data set as to save computational costs.
Now, we define the normalized fluctuation as

\begin{equation}
    %\hat{x}_i = \frac{x_i^\prime}{\underset{j\in X}{max}(|x_j^\prime|)}
    \hat{x}_{ij} = \frac{x_{ij}^\prime}{max_{k}(|x_{ik}^\prime|)}.
\end{equation}

The weighted average effect of $x_i$ on $y$ can now be quantified in a similar way as in \citeA{beucler2024-climate-inv} in a vector $\mathbf{S}$, with

\begin{equation}
    \label{eq:S_j}
    S_i = \langle \hat{x}_{ij}\cdot shap(x_{ij},y)\rangle_{j}.
\end{equation}

A positive $S_i$ expresses an increasing/decreasing $y$ for an increasing/decreasing $x_i$ independently of other values, and for a negative $S_i$ we have the opposite effect.
For a multi-output model this vector $\mathbf{S}$ becomes a 2D matrix $S_{ij}$ quantifying the influence of the $i$th input on the $j$th output. We will refer to the SHAP values obtained by this method as weighted SHAP values from now on.

\begin{figure}[tbh]
    \centering
    \includegraphics{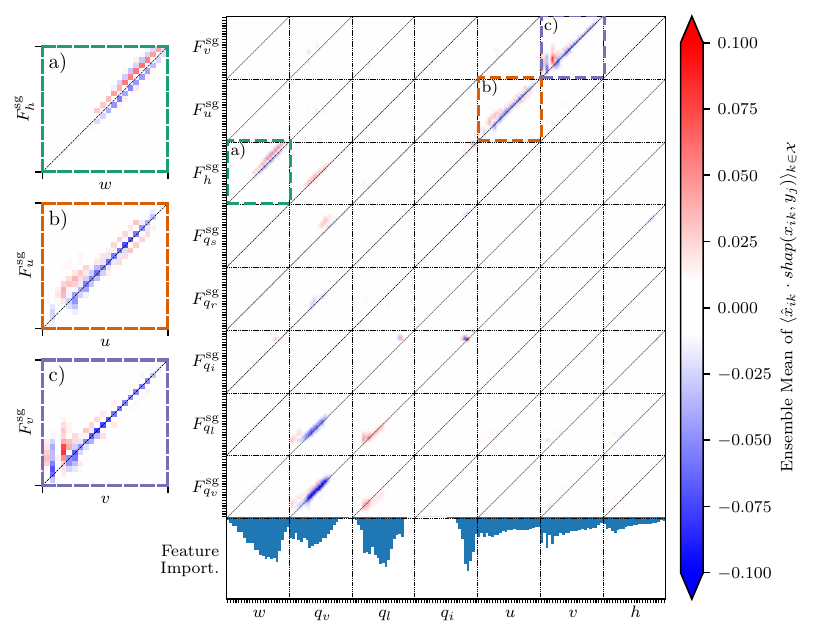}
    \caption{Ensemble mean of weighted SHAP values aggregated according to \cref{eq:S_j} for the U-Net. The variables $q_r,\allowbreak{}q_s$ were ablated. The height level for each variable is increasing from left to right / from bottom to top. The feature importance depicted in the lower part of the figure shows the mean absolute SHAP values averaged over all target fluxes. Insets a), b), and c) show a zoom into the plot for three specific variable pairs, the colors indicate which inset corresponds to which part of the large plot.}
    \label{fig:EnsembleMeanShapRhoFluctWoQrQs}
\end{figure}

Applying this method to the trained U-Net model gives the matrix visualized in \Cref{fig:EnsembleMeanShapRhoFluctWoQrQs}.
We see many interpretable, vertically local influences (main diagonal patterns) in this figure, for example, controlling for $q_l$, there is a mainly negative influence of specific humidity $q_v$ on $F^\mathrm{sg}_{q_v}$/$F^\mathrm{sg}_{q_l}$ visible.
As previously observed by \citeA{beucler_et_al_ALinearResponse}, this vertically local drying effect is plausibly related to the entrainment of water vapor into convective plumes and its subsequent downwards advection \cite{beucler_et_al_ALinearResponse}.
Moreover, an increase in water vapor also increases the moisture gradient to the environmental air and leads to the entrainment of drier air.
The local drying effect is seen for levels in the lower to middle troposphere, approximately at \qty{700}{\metre} to \qty{5}{\kilo\metre}.
Furthermore, we see a slightly positive impact and moistening flux of the lower model levels on higher levels.
This is indicative of the decrease in air density for increased water vapor content and the decreased lapse rate for buoyant air parcels (and therefore higher convective instability).
For cloud liquid water $q_l$ the opposite effect can be observed on the convective subgrid fluxes of $q_v$/$q_l$.
This learned correlation can be understood by looking at the condensation process of water vapour.
When water condenses in an atmospheric grid cell, latent heat is released and the air becomes more buoyant.
This in turn can lead to more condensation and therefore to moisture convergence in the area and cloud formation.
Furthermore, more liquid water can lead to precipitation.
The evaporation of falling raindrops can consequently lead to an increase in local humidity, especially if the layers below are far from saturation.
Finally, hygroscopic effects could play a role as cloud droplets can act as condensation nuclei, attracting more water vapor and leading to cloud growth.

A direct comparison with the linearized response functions from \citeA{RN108} and \citeA{kuang_linear_stability_part1} is difficult as we use different variables and a dataset from a non-idealized simulation (e.g., no aquaplanet configuration, active diurnal cycle, and spherical simulation domain).
Nevertheless, for the influence of water vapor on the subgrid flux of water vapor and cloud liquid water we see similarities to the response of $Q_2$ (apparent moistening) to the total nonprecipitating water mixing ratio in \citeA{RN108}.
For both analysis methods a vertically local negative influence is visible.
In the study \citeA{kuang_linear_stability_part1} this response is similarly traced back to the impact of relative humidity on the specific humidity tendency.
Furthermore, we also observe a positive convective heating response to an increase in moisture (influence of $q_v$ on $F^{sg}_h$) as shown in both studies, although more local in this study as opposed to a non-local heating of higher layers in response to a moistening lower troposphere.

Apart from that, the main visible signatures are visualized in the insets of \Cref{fig:EnsembleMeanShapRhoFluctWoQrQs}.
Inset a) shows the influence of $w$ on $F^\mathrm{sg}_h$.
The main pattern is in the upper layers where we can see primarily a positive super- and negative sub-diagonal ($S_{ij}$ with $j=i-1$ and $j=i+1$, respectively).
This means that cells with a high vertical velocity have a positive influence on the subgrid flux in the cell above them and a negative influence below them respectively.
Considering that mesoscale convergence and large scale ascent can initiate/enforce convective cells \cite{KALTHOFF2009680}, this seems reasonable.
Below the convective region, the atmospheric column becomes more stably stratified, explaining the negative sub-diagonal of the figure.
In Inset b), a negative diagonal pattern with some positive signatures above can be observed.
Consequently, high horizontal wind speeds imply a positive horizontal momentum flux to higher levels.
This signifies that the U-Net has learned a downgradient diffusive momentum flux parameterization.
We also see a positive pattern in the sub-diagonal for higher levels looking at subplot b).
Vertical wind shear has been found to be an essential ingredient for long-lived and well-organized convective storm cells \cite{rotunno1988,DOSWELL2003117,ASimpleModelConvectiveSystems}.
A very similar pattern can be observed in Inset c), the main difference is that for lower levels there are a few vertically non-local transport signatures.
These patterns are consistent (with relative standard deviations of max $\sim\qty{10}{\percent}$) over different realizations of $\mathcal{X}$ so that the result here seems not to be dependent on the set $\mathcal{X}$.

\begin{figure}[tbh]
    \centering
    \includegraphics{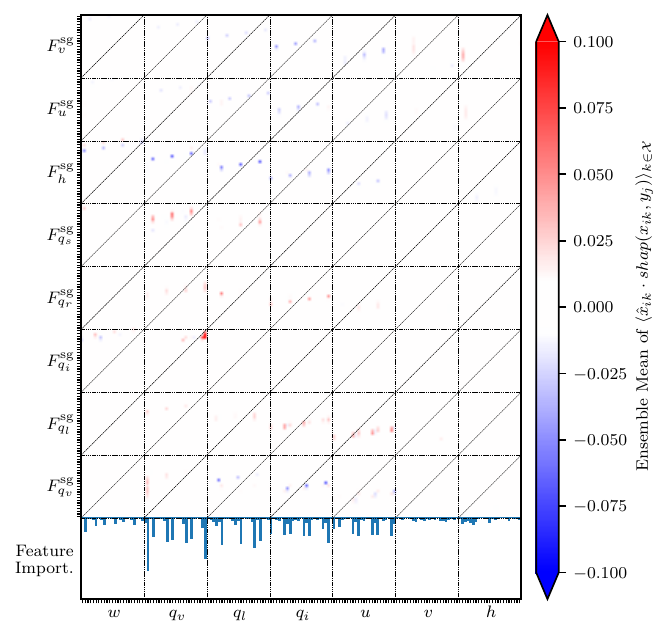}
    \caption{Ensemble mean of weighted SHAP values aggregated according to \cref{eq:S_j} for the GBT model. The variables $q_r,\allowbreak{}q_s$ were ablated. The feature importance depicted in the lower part of the figure shows the mean absolute SHAP values averaged over all target fluxes.}
    \label{fig:EnsembleMeanShapGBTRhoFluctWoQrQs}
\end{figure}

As a comparison, the corresponding weighted SHAP values for the GBT are displayed in \Cref{fig:EnsembleMeanShapGBTRhoFluctWoQrQs}.
First, the GBT feature importances have a much less regular pattern and look more ``randomly'' distributed.
These patterns show a less coherent picture and are not so easily interpretable.
Looking at the aggregated feature importance, both models weigh the liquid/ice water static energy the least.
The GBT model weighs the specific humidity higher in its predictions with an aggregated importance of \qty{29}{\percent} compared to the U-Net with \qty{14}{\percent}.
As most important features for the U-Net, on the other hand, we have the vertical velocity $w$ and cloud water content $q_l$.
These two variables are also part of the condition formulated in \cref{eq:ConvectiveConditions} for convective conditions in a grid cell.
Therefore, it is reasonable that the network learns to pay attention to these inputs.

Since the weighted SHAP values displayed in \Cref{fig:EnsembleMeanShapRhoFluctWoQrQs} consistently show vastly different patterns than in \Cref{fig:EnsembleMeanShapGBTRhoFluctWoQrQs}, we used the same method for the RF as well and got a similar picture to what is displayed here for the GBT.
In order to rule out a dependence of the obtained results on the Shapley value approximation method, we also used the \texttt{Kernel\-Explainer} \cite{NIPS2017_7062} as an alternative to the \texttt{Tree\-Explainer}.
The resulting weighted SHAP values have almost the same form as for the \texttt{Tree\-Explainer} class, emphasizing that our results are independent of the explanation method.
We also looked at the standard deviations of all weighted SHAP value plots and observed that the uncertainty is very low compared to the mean values shown (The maximum deviation is $0.02$, and \qty{99}{\percent} of the standard deviation values are below $0.002$), further demonstrating that those interpretation statistics are stable across samples.

For the data in \Cref{fig:EnsembleMeanShapRhoFluctWoQrQs}, these values are $0.02$ and $0.004$, respectively.
Looking at the scales in both figures, these uncertainties are very small.

Overall, this indicates that although the predictive performance of the GBT is comparable to that of the U-Net, it relies on very different statistical patterns in the data.
These patterns are more non-local and mostly unphysical so that the resulting model is expected to have less skill in extrapolating outside its training domain.

\subsection{Online Stability Tests}\label{sec:online_stability_tests}
In this section we will test the U-Nets ability to run stable in a coupled setting and consequently test their (global) extrapolation capabilities.
We do not perform an offline extrapolation test with another data set since the hypothesized non-causality of the full U-Net would not show any negative impact in this test.
For this reason we decided to couple the developed parameterizations back to the host (ICON) model and thus have a stronger generalization test.
We first couple both the ablated (without precipitation tracer inputs) and the full U-Net to the ICON model and observe that the ablated U-Net shows improved stability compared to the full U-Net, when coupled globally.
We also find that the ablated U-Net gives improved extreme precipitation predictions as opposed to the full U-Net, which fails to predict the precipitation distribution accurately.

Coupling data-driven parameterizations to GCMs is typically intricate and the stability of the developed schemes is very sensitive to e.g., changes in the training data set \cite{gmd-13-2185-2020} or the inclusion of variables on specific levels and the choice of the loss function \cite{RN107,RN108}.
Trial and error is often used to find stable schemes among the offline trained parameterizations \cite{RN125}.
Stability issues of coupled models have been observed, even for idealized setups such as aquaplanet simulations \cite{RN5,RN19,RN17,RN143}.
Other studies, in which coupled ML schemes have used more realistic setups, were trained and coupled with superparameterized GCMs \cite{iglesias-suarez_causal_nns,RN125,han-an-ensemble-NN-moist-physics}.
A technical advantage of training on these datasets is that a clear scale separation is artificially introduced and therefore the training targets for the ML algorithms are well defined.
On the other hand, this scale separation influences the emergent dynamics and the embedded SRMs are themselves idealized as they are 2D models with a limited extent \cite{pritchard-restricting-horizontal-scales-MJO,RN143}.

Introducing a new parameterization into a GCM typically requires a retuning of the host model to e.g., adjust for current compensating biases in the interplay of various parameterization schemes \cite{2023arXiv230408063G}.
There are potentially many feedbacks when coupling a new scheme to the GCM which can quickly lead to unstable configurations or incorrect results.
Furthermore, because there are considerable design differences between storm-resolving and coarse-resolution global climate models \cite{Satoh2019-global-cloud-resolving-models}, there could be distributional shifts between both types of model classes.
Substantial distributional shifts have already been observed within the class of storm-resolving models \cite{Mooers2023}, so that ML parameterizations trained on data from a different storm-resolving model cannot be expected to learn the same relations.
Also, by coarse-graining high-resolution fields, disturbances which can be represented on the coarse grid but not accurately advected by the coarse model can be introduced as noted by \citeA{watt-meyer-NN-param-realistic-SRMs}.
To tackle this problem and to keep the coarse dynamics close to the coarsened high-resolution state, they nudged the coarse simulation to a coarse-grained high-resolution reference state continuously and achieved stable coupled runs (with ML-predicted tendencies for heat and moisture) for about 35 days with realistic boundary conditions.

Because of these issues and limitations we do not expect our models to show accurate online performance without some further modifications.
Nevertheless, we tried to couple the U-Net models to the ICON model to test their stability and therefore our hypothesis about the extrapolation capabilities of the full and ablated U-Net.
For this coupling we used the FTorch library \cite{FTorch} to load our models within ICON and to run them in inference mode during the time integration.
Before the actual coupling we added a preprocess/postprocess layer to both NNs which normalize all the input variables to zero mean and unit variance and apply a corresponding inverse transformation for the output variables.

To test the stability of our developed ML parameterizations we created four different ICON configurations:
\begin{enumerate}
    \item Ablated U-Net applied for all longitudes and latitudes
    \item Full U-Net applied for all longitudes and latitudes
    \item Ablated U-Net applied for all longitudes and only tropical latitudes
    \item Full U-Net applied for all longitudes and only tropical latitudes
\end{enumerate}
For configuration 1 and 2 the convection schemes have to extrapolate substantially as e.g., temperature, humidity, and also wind patterns differ considerably in the extratropics.
Configurations 3 and 4 are applied closer to their training data set domain, i.e. the tropics.
We apply the U-Nets between the Tropic of Capricorn (\qty{23.43616}{\degree}S) and the Tropic of Cancer (\qty{23.43616}{\degree}N) while the training domain is approximately defined between \qty{10}{\degree}S and \qty{20}{\degree}N as shown in \Cref{fig:AverageConvectiveCellsAcrossNarval}.
Outside of the tropics the conventional mass-flux convection scheme is applied for these two configurations (3/4).
For all coupled simulations (and the reference simulations), we use ICON in its version 2.6.4, with an \texttt{R2B5} ($\Delta x\approx\qty{80}{\kilo\metre}$) horizontal grid and 47 vertical layers.
Parameterized processes include radiation, cloud microphysics, orographic and non-orographic gravity wave drag, turbulence, and (ML-based) convection.

\begin{figure}[tbh]
    \centering
    \includegraphics{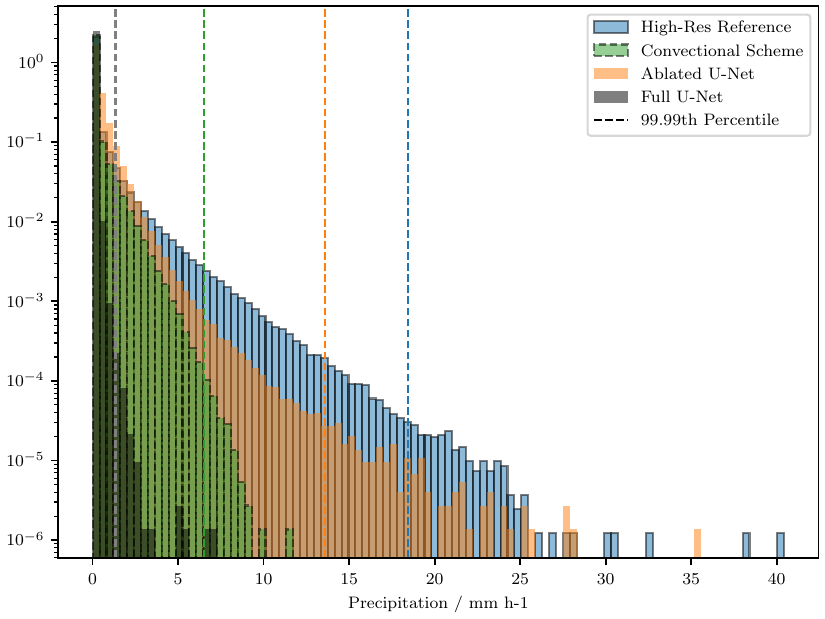}
    \caption{The precipitation distributions of the first two weeks over the tropics for the three simulations starting on 01.02.1979 for  the full U-Net (configuration 4) in grey/dark green, the conventional cumulus scheme in green, and the ablated U-Net (configuration 3) in orange. Also, the precipitation distribution for the high-resolution data set (NARVAL) is displayed in blue. The $99.99$th percentiles of each data set are marked by dashed lines in the corresponding color.}
    \label{fig:PrecipDistributionOnline}
\end{figure}

We initialized a simulation from interpolated IFS (Integrated Forecasting System) analysis data for the 01.01.1979 and ran the ICON model for one month.
After this initialization phase we wrote out initial conditions for each day at 0000\,\unit{UTC}.
With these initial conditions we started ten new runs with a length of half a year for each model configuration (from the 01.02.1979, 02.02.1979, \dots, 10.02.1979) to test the stability of the ML schemes.
For columns with the ML scheme activated we applied the tendencies for heat, moisture, zonal and meridional wind which are derived by taking divergence of the ML-predicted fluxes instead of the ones derived by the conventional mass-flux scheme.
Everywhere else, only the conventional convection parameterization of the ICON model was applied.
No switch condition for the activation of our ML scheme was needed as we chose to add \qty{10}{\percent} of non-convective columns to the training data set, as explained in \cref{sec:data_and_preprocessing}, so that the U-Net learned when not to predict any convective fluxes.
An alternative option would be to use the trigger condition from the conventional cumulus scheme, where convection is triggered for columns with moisture convergence, and some thresholds regarding humidity and buoyancy must be met \cite{moebis2012-factors-controlling-itcz-aquaplanet}.
We decided to not use such condition here but we could explore such methods in future work.

A first result of the online simulations is shown for the probability density function of precipitation in \Cref{fig:PrecipDistributionOnline} and for the spatial distribution of mean precipitation in Figure \ref{fig:PrecipSpatialMean4MonthCoupledSims}.
In \Cref{fig:PrecipDistributionOnline}, the distribution of precipitation over the first two weeks of simulation over the tropics is displayed for the setup with the conventional cumulus scheme,  the ablated U-Net (configuration 3), and the full U-Net (configuration 4).
For both, configuration 3 and configuration 4, we set values of negative precipitation to zero.
In future work this could be avoided by using an activation function with a non-negative codomain, like the \texttt{relu}-function, for precipitation.
For the simulations shown here we set the large-scale precipitation to zero as said in \cref{sec:data_and_preprocessing}.

The spatial distribution (monthly means) of precipitation over the region where we have a high-resolution reference can be seen in Figure \ref{fig:PrecipSpatialMean4MonthCoupledSims} in the supplementary information.
The spatial mean precipitation patterns show that the coupled ablated U-Net results in a much more reasonable spatial distribution of precipitation than the full U-Net which heavily underestimates the mean precipitation.
Compared to the high-resolution reference, the ablated U-Net produces a spatially more uniform precipitation distribution and has regions with too high mean precipitation.
The conventional scheme shows a heavy land bias for the mean precipitation and shows too low precipitation values.

\Cref{fig:PrecipDistributionOnline} demonstrates the potential and added value of ML parameterizations as the precipitation distribution for the coarse model coupled with the ablated U-Net is much closer to the high-resolution (NARVAL) distribution than the reference simulation.
For the full U-Net (configuration 4) we see an opposite effect: the distribution does show even less extreme values than the simulation with the conventional cumulus convection parameterization.
This shows that the full U-Net, which heavily relies on the precipitation tracers (see Figures \ref{fig:ShapImportanceComparison} and \ref{fig:EnsembleMeanShapRhoFluct}), struggles to show good online performance.
The reason lies in the hypothesized non-causal relations to the mentioned precipitation tracers.
In coarse-grained (offline) data, precipitation is highly informative about convective events and further precipitation due to convective memory but as soon as the parameterization is coupled, the scheme struggles as the ML model itself has to predict some convective fluxes and precipitation in the first place.

The values for the 99.99th percentile further show the increased ability of the ablated U-Net to predict precipitation extremes more accurately and therefore the potential to reduce the common problem of GCMs to predict these extremes accurately \cite{RN158}.
These percentile values are \qty{18.44}{\milli\metre\per\hour} for the NARVAL data, \qty{13.07}{\milli\metre\per\hour} for the ablated U-Net, \qty{6.67}{\milli\metre\per\hour} for the reference simulation, and only \qty{1.34}{\milli\metre\per\hour} for the full U-Net.

\begin{figure}[h!]
    \centering
    \includegraphics{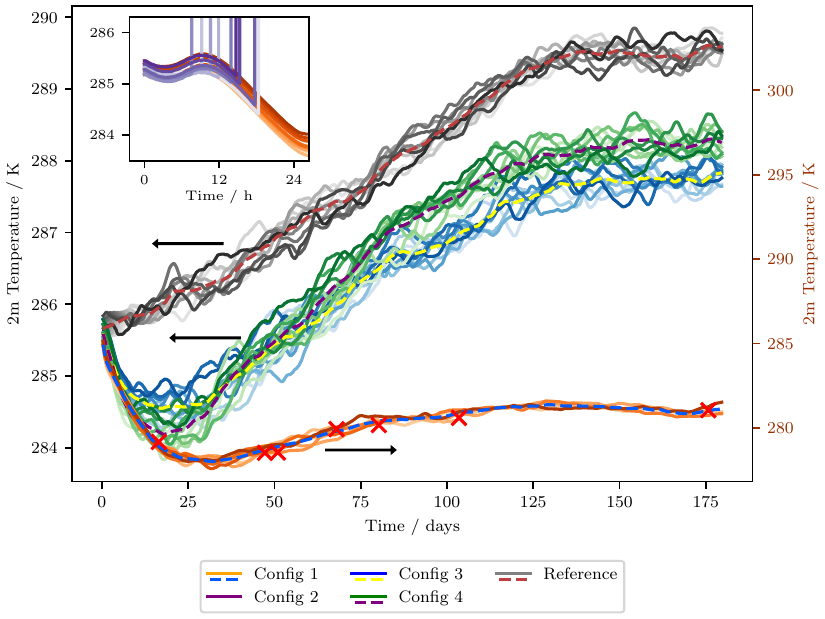}
    \caption{The stability of the ablated vs. the full U-Net in form of a time series of the global mean air temperature on \qty{2}{\metre} height over 180 days. For each defined configuration, the ten realizations are drawn in orange, purple, blue and green colors, respectively. Solely for the full U-Net coupled globally (Config 1), a second y-axis (also in orange) on the right side of the plot is introduced as this simulations shows a much higher reduction in 2m Temperature. To make this clearer, arrows are indicating the corresponding y-axis for each ensemble. An inset provides a close-up of the first 24 hours of the dynamics of configurations 1 and 2: the simulations with the full U-Net quickly become unstable. The data displayed in the inset has been saved with an output frequency of six minutes as opposed to the more stable simulations with an output frequency of six hours in the main plot. For all of the data except the inset, a rolling mean over \qty{24}{\hour} was applied. Additionally, multi-model means over configurations 1, 3, 4, and the reference ensemble, respectively, are drawn as light green/yellow/violet/red-brown dashed lines. These colors are chosen as the complementary colors of the respective ensemble members and are marked in the legend as the second lower dashed line for each ensemble. For configuration 1, model blow-ups are marked by red crosses as to not obscure the other lines.}
    \label{fig:2mTemperatureOnlineEnsembleAlllonlat}
\end{figure}

Looking at the stability of the coupled simulations, \Cref{fig:2mTemperatureOnlineEnsembleAlllonlat} displays the global mean surface temperature of all simulations of configuration 3 and 4 for 180 days.
We can see that all simulations of configuration 3/4 are stable for the displayed period while the simulations with the full U-Net applied globally (config 2) very quickly become unstable, after about 6 to 18 hours.
Configuration 1 (ablated U-Net coupled globally) is stable for the first day and simulations diverge only over the course of half a year as it can be seen for the orange lines in \Cref{fig:2mTemperatureOnlineEnsembleAlllonlat}.
The simulations are stable for about 115 days on average with two simulations from these configurations staying stable for all 180 days.
For the fully stable simulations, the surface temperature initially drops by about \qty{1}{\kelvin} and then increases again to a higher value than the initial temperature.
By looking at the full 180 days of time integration, the temperature for configuration 3/4 seems to equilibrate at about \qty{287.8}{\kelvin}/\allowbreak{}\qty{288.2}{\kelvin} ($\sim$\qty{14.7}{\celsius}/\allowbreak{}\qty{15}{\celsius}) as seen in the figure.
This is not unrealistic but the main point of this figure is to show the coupled stability for multiple months which is already three times the length of the training data set (two months).
The global mean temperature of configuration 4 shows a similar trend compared to configuration 3 but becomes stable at slightly higher temperatures.
The reason could lie in the fact that convection is much more infrequent for the full U-Net configuration as it can be seen in \Cref{fig:PrecipDistributionOnline} and heat is therefore transported less efficiently to higher levels.
Comparing these two fully stable simulations to the reference simulation mean in grey, we can see that there is an initialization shock (the mentioned initial temperature drop) \cite{bretherton22-CorrectingCoarse}.
After this shock, the seasonal variation appears very similar in both magnitude and phase to the reference simulations.
The initial shock indicates the off-set, that would have to be addressed by tuning, or nudging as in \citeA{watt-meyer-NN-param-realistic-SRMs}, as described earlier.

As all ensemble members of configuration 2 quickly diverge (as opposed to configuration 1, which is stable for minimally 16.5 days) we conclude that our hypothesis, that the full U-Net learned non-causal relationships, gains more support.
However, the ablated U-Net configuration, does not guarantee stability when coupled globally.

\begin{figure}[tbh]
    \centering
    \includegraphics[width=1\textwidth]{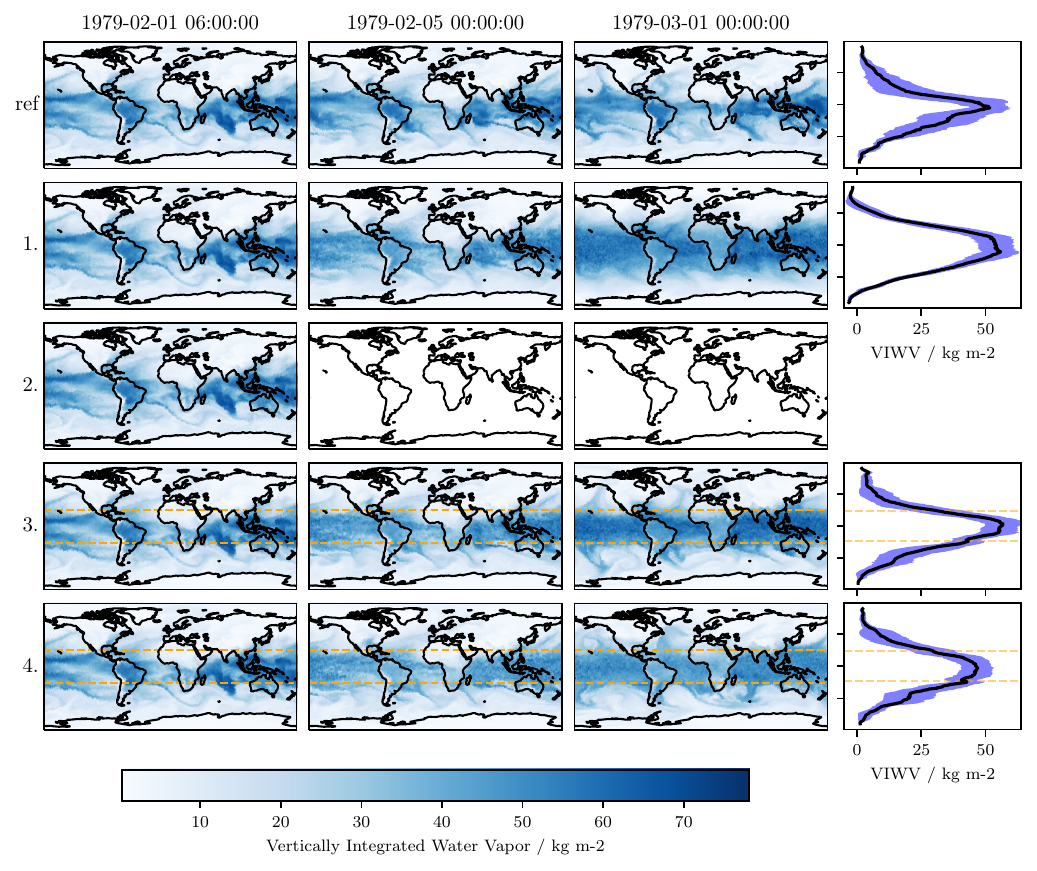}
    \caption{The vertically integrated water vapor for three simulation snapshot with convection parameterized with the conventional physical cumulus convection scheme of the ICON model as a reference (ref), 1.) by the ablated U-Net, 2.) by the full U-Net, 3.) by the ablated U-Net applied only in the tropics, 4.) by the full U-Net applied only in the tropics. For row 3.) and 4.) the domain where the ML schemes are applied are marked by orange dashed lines. The last column shows the zonal mean and standard deviation of the vertically integrated Water Vapor (VIWV) for the last shown date (1979-03-01) of every configuration except the unstable one. The y-axis corresponds here to the latitudes of the corresponding row.}
    \label{fig:VerticallyIntegratedWaterVaporOnline}
\end{figure}

To have a closer look at the dynamics we show the vertically integrated water vapor in \Cref{fig:VerticallyIntegratedWaterVaporOnline}.
A reference simulation with the conventional cumulus convection scheme is shown in the top row and the other rows are marked by their configuration number as defined above.

For the coupled full U-Net applied at all latitudes/longitudes we can only see one snapshot after six hours in \Cref{fig:VerticallyIntegratedWaterVaporOnline} because for the other dates the simulation has already diverged.
For the snapshots after 4 days of simulation the structures with the ML parameterizations still look close to the reference simulation but there can already be seen some blurring effects in the tropics, especially over the ocean (e.g., over the Pacific).
After a month of simulation configurations 1, 3, and 4 lost most of the structure in the tropics and instead there is a homogeneous high water vapor accumulation over these latitudes.
This blurring effect is also displayed in the zonal mean and standard deviation plots in the last column.
Especially for the ablated U-Net coupled globally (configuration 1), the standard deviation in the extratropics is very low.
Furthermore, it is visible that for the ML coupled simulations, the mean water vapor path has a flatter peak compared to the reference and for the coupled full U-Net, the water vapor path has, additionally, a smaller magnitude in general.
Note that for the ablated U-Net coupled globally (configuration 1), \Cref{fig:VerticallyIntegratedWaterVaporOnline} shows that the zonal mean water vapor path is less than zero for very high latitudes.
This demonstrates the ML models failure to extrapolate to these latitudes, although, as most of the extratropical values still look reasonable and this configuration is stable compared to the full U-Net, this degree of extrapolation could also be considered unanticipated.

The blurring problem is a very common one for data-driven atmospheric models and can be related to the fact that ML models minimize a deterministic error and tend to predict some mean state rather than, possibly a more realistic, extreme state \cite{rasp2023-weatherbench2}.
While this explanation cannot directly be transferred for the smoothing we see here, as we did not develop a fully data-driven atmospheric model, the used ML models are also incentivized to predict mean fluxes due to the used deterministic RMSE.

A similar effect has been observed by \citeA{kwa2023-MlClimateModelCorrections}, by applying ML corrections to their coarse GCM they observed a reduction in tropical variability of precipitation.
Alternatively, the existence of the observed blurring could be caused by the comparably low accuracy of U-Net at lower levels (see \Cref{fig:R2HeightProfiles}) or the U-Net's failure to represent convection over steep orography.
Outside of the tropics, where the ML parameterization is not applied, there are still some structures, e.g., atmospheric rivers, visible in the configurations 3 and 4.

As we said before, there are many challenges to coupling an offline trained parameterization to a GCM and the results in \Cref{fig:VerticallyIntegratedWaterVaporOnline} show that, although many simulations run stably for a long time, there is still much room to improve the ML algorithms.
Nevertheless, we were able to test the stability of our developed data-driven schemes.
Both the ablated and the full U-Net support stable simulations when coupled only inside tropical latitudes.
However, coupling the full U-Net, for which we hypothesized non-causal relations (see Figure \ref{fig:EnsembleMeanShapRhoFluct}), leads to model blow-ups rather quickly when coupled globally, outside the training domain.

\section{Conclusions and Discussion} \label{sec:discussion_conclusions}
In order to develop an ML-based parameterization for convection we first filtered, processed, and coarse-grained data from high-resolution  simulations with explicit convection.
To separate convection from other processes, we used a filtering method for convective conditions.
That ensures that the ML models learn mostly convective fluxes.
We then coarse-grained the high-resolution data to the target resolution and calculated the subgrid fluxes of the needed output quantities.
The coarse-graining was performed without neglecting horizontal density fluctuations since we used data from a model with terrain following coordinates and the irreducible error increases if the model does not have the necessary input information.
For the vertical coarse-graining we had to neglect some columns from the data set with especially steep orography.
However, there are still many columns over heterogeneous terrain available and most trained models are able to run stable online.
Nevertheless, future work could target including also these column and therefore profit from a orographically more diverse data set.

We found that the U-Net architecture is a very suitable machine learning model to parameterize convective subgrid fluxes, which is naturally a multiscale process.
The U-Net outperformed other deep learning models by only a small margin judging by the $R^2$ metric.
However, comparing the offline performance over a broad range of parameters, the error of the U-Net was consistently lower than the error of MLP, CNN, and ResNet architectures (\Cref{fig:NNTestLossCurves}), this showed the structural advantage of the U-Net compared to the other models.
A comparatively lower $R^2$ is achieved by most non-deep-learning models except for the Gradient Boosting Trees model.
The linear models show a higher performance compared to the random forest and extra tree regression model.
This could be related to the missing extrapolation capability of these tree based models, the effective feature selection of these regularized linear models, or, possibly, due to too heavy tuning to tropical convection.
We will have to conduct more experiments in future research to train and test these models globally.
Based on our offline evaluation we cannot claim that the tree-based models are not able to perform well online, therefore we plan to explore the online performance of the tree-based models by coupling them to ICON as well.
The coupling of tree-based models to a GCM has been done successfully before by, e.g., \citeA{RN17,RN145} (although, in idealized aquaplanet settings).
The GBT model had a coefficient of determination of $R^2\approx0.84$ compared to the U-Net with $R^2\approx0.90$.
Nonetheless, in a direct comparison between GBT and U-Net, the best performing non-deep learning and deep learning model, the U-Net had an advantage in almost all aspects.
An exception to this is shown in \Cref{fig:ConvectiveCellProfile} by the $R^2$ value for a few levels for ice, snow, and cloud water tracers.
For snow and ice these exceptions occurred in the lower levels and for cloud liquid water mainly in the higher ones, where the respective tracer species are rarely observed / have a very low concentration.
This demonstrates the advantage of the lower complexity tree-based method for sparse data or rather for regions where an interpolation based on few relevant samples is needed.
For the other levels and also for the predicted 2D fields, such as convective precipitation, we noticed a clear benefit of using the U-Net architecture.
We do not claim exhaustiveness in the choice of ML models/NN architectures, the parameterization could profit from the combination of specific architectures benchmarked here, such as ResNets and CNNs, or other more advanced model such as recurrent NNs or Transformers (with the height as time/sequence dimension).

While the U-Net shows a high skill in parameterizing multiscale convection, we did not empirically test the multiscale representation of the NNs.
Future research could target testing these multiscale properties by e.g., ablating the most compressed layers and looking at the decrease in accuracy for deep convection or testing the ability of the model to work on scaled in/outputs.
Furthermore, other modifications, like dilated convolutions \cite{2015arXiv151107122Y}, could be tried to enhance the multiscale processing of the U-Net.

To get some insight into what the model exactly learned during training we applied the SHAP framework and first calculated feature importances.
These revealed that the U-Net model focuses strongly on the precipitating tracer species rain and snow as input variables.
Here, the SHAP values exposed that the model learned non-causal relations between convective subgrid fluxes and convective precipitation.
This was also seen in the figure showing the weighted SHAP values (Figure \ref{fig:EnsembleMeanShapRhoFluct}), as particularly the rain tracers showed heavy non-local influences on subgrid fluxes for liquid/ice water static energy, rain, cloud liquid, and water vapor tracers.
For comparison, the weighted SHAP values for the MLP model can be seen in Figure \ref{fig:EnsembleMeanShapRhoFluctMLP}.
Similar non-causal connections to precipitating tracer species can be observed in that figure and, in fact, we found that for all deep learning models with a full input, the precipitating tracer species show the highest (shap value-based) feature importance assignment.
As a result we performed the same analysis on an ablated model without water species. 
A potential solution to be investigated in a future study would be to restrict the model to learn causal relationships as in \citeA{iglesias-suarez_causal_nns}.
Another approach to improve the predictions of subgrid momentum fluxes specifically would be to model the degree of small scale convective organization \cite{RN119}.
For higher stability in coupled simulations of the developed ML-based multi scale parameterization to a GCM it will be advantageous to use a global training data set.
Convectively active regions in the extratropics would be especially important to include, e.g., regions where frontal systems and extratropical cyclones are common, extratropical monsoon regions, or locations with marine stratocumulus clouds.
Furthermore, it would be important that ML models learn the distributions corresponding to, e.g., the arctic climates so that out-of-distribution predictions can be avoided in high latitudes.

By looking at the weighted SHAP values we found that the ablated version of the U-Net was more physical and learned physically explainable connections between coarse-scale variables and subgrid fluxes.
For example, there were patterns indicating local upwards transport of horizontal momentum and energy, moisture convergence, and the interaction between wind shear and mesoscale convective systems.
This strengthens trust in the model as it can be expected to extrapolate better to data outside its training domain.
However, many interpretations of the weighted SHAP value matrices, besides some objective features like locality, are rather subjective (e.g., mesocale convergence) and should be generally regarded as one out of many tools to build up trust in the models.
The weighted SHAP values for the GBT model were not physically interpretable as they showed very scattered results and close to no coherent patterns.
We applied a different explainer class to test the robustness of this outcome and saw consistent results.
To investigate this further, we did the same analysis for the Random Forest as this model has been used in other studies before.
Here, the weighted SHAP values were similarly scattered as for the GBT model.
This result shows that seemingly well performing models (judging by e.g., $R^2$) can in fact rely on non-causal correlations in the data, achieving good results for the ``wrong reasons''.
Therefore, these models are most likely not suited for the coupling to a GCM.
The emergence of these non-causal relationships and possible methods of prevention, besides ablation, should be investigated further in future research.

In the section on online stability tests we coupled the ablated and full U-Net to the ICON model and showed that, when coupled globally, the hypothesized non-causal connections indeed lead to instability within a day for the full U-Net; as opposed to the ablated U-Net which support stable simulations for minimally 16 days (and on average, 115 days).
For the ablated U-Net (and both U-Net parameterizations applied only in the tropics) we found stable simulations for at least 180 days.
By coupling the ablated U-Net to the ICON model, we could show that the ML model is able to predict precipitation extremes more accurately online (see \Cref{fig:PrecipDistributionOnline}) in contrast to the conventional parameterization and the full U-Net.
The stable simulations are showing e.g., smoothing biases already after some weeks.
Tracing back the specific output variables responsible for this smoothing bias would be significant to understanding and minimizing this effect in future research.
An approach using a stochastic ML parameterization could mitigate the smoothing bias, possibly, related to the usage of the RMSE mentioned in \cref{sec:online_stability_tests}.
However, we did not expect perfect results because of distributional shifts between the training data set and the variable states of the coarse simulation.
Furthermore, as our process separation is not perfect and at least some momentum fluxes from gravity waves will have an impact on the dynamics, we will do some further tests in the future e.g., without a parameterization for non-orographic gravity wave drag.
Another possible approach for future research would be to build a combined parameterization for convection and microphysics to more accurately represent their interaction and the influence of convective updrafts on microphysics.
For further improvement of the coupled model results it might be necessary to train the models on a global data set, use climate-invariant variables \cite{beucler2024-climate-inv}, or work on more physically constrained architectures \cite{beucler2023-ml-clouds-climate}.
With more physically constrained and robust ML parameterizations, an extensive validation against a range of climatic conditions to ensure that any improvements in parameterization translate to more accurate climate representations would be necessary.

Our study leads to the conclusion that interpretability/explainability of ML algorithms is important to investigate potentially non-physical mechanisms.
Furthermore, we conclude that the U-Net is the best choice of the examined model classes as it is very accurate, not too complex, and its predictions can be explained physically after domain knowledge was applied to ablate spurious correlations.
This advantage over other ML-model classes likely comes from the ability of the U-Net to capture multiscale phenomena like convection.
In the future, we will expand our work by training ML models on global high-resolution data for which we ensure that input variables and fluxes are output after the dynamical core or respectively, after parameterizations for processes which are neither resolved for the high-resolution simulation nor the coarse scale, e.g., radiation.
By doing this, we will avoid distributional shifts between the coarse-grained data set and the coarse simulations.

\clearpage
\appendix

\section*{Open Research}
The code is published under \url{https://github.com/EyringMLClimateGroup/heuer23_ml_convection_parameterization} and preserved \cite{helgehr_2024_12773936_article}.
The simulation data used to train and evaluate the machine learning algorithms amounts
to several TB and can be reconstructed with the scripts provided in the GitHub repository.
Access to the NARVAL data set was provided by the German Climate Computing Center (DKRZ)
The software code for the ICON model is available from \url{https://code.mpimet.mpg.de/projects/iconpublic}.

\acknowledgments
Funding for this study was provided by the European Research Council (ERC) Synergy Grant ``Understanding and Modelling the Earth System with Machine Learning (USMILE)'' under the Horizon 2020 research and innovation programme (Grant agreement No. 855187).
This work used resources of the Deutsches Klimarechenzentrum (DKRZ) granted by its Scientific Steering Committee (WLA) under project ID bd1179.
The authors gratefully acknowledge the Earth System Modelling Project (ESM) for funding this work by providing computing time on the ESM partition of the supercomputer JUWELS \cite{JUWELS} at the Jülich Supercomputing Centre (JSC).
Furthermore, we thank the authors of \citeA{RN57} for creating and providing the high-resolution simulations of the tropical Atlantic used in this study.

\bibliography{CombinedLib}%,SecondLibrary}% don't specify the file extension

\section*{References from the Supporting Information}

\dummylabel{sec:appendix_add_figures}{S1}\noindent\textbf{Section S1}

\dummylabel{sec:appendix_nondl_info}{S2}\noindent\textbf{Section S2}

\dummylabel{sec:appendix_hpo}{S3}\noindent\textbf{Section S3}

\dummylabel{fig:SubgridFluxScatter_qlqiqrqs}{S1}\noindent\textbf{Figure S1}

\dummylabel{fig:PrecipitationDistribution}{S2}\noindent\textbf{Figure S2}

\dummylabel{fig:NormRmseSpatialComparison}{S3}\noindent\textbf{Figure S3}

\dummylabel{fig:EnsembleMeanShapRhoFluct}{S4}\noindent\textbf{Figure S4}

\dummylabel{fig:EnsembleMeanShapRhoFluctMLP}{S5}\noindent\textbf{Figure S5}

\dummylabel{fig:ComplexityPerformance}{S6}\noindent\textbf{Figure S6}

\dummylabel{fig:PrecipSpatialMean4MonthCoupledSims}{S7}\noindent\textbf{Figure S7}

\dummylabel{fig:HpoVisualization}{S8}\noindent\textbf{Figure S8}

\dummylabel{tab:net_num_params}{S1}\noindent\textbf{Table S1}

\dummylabel{tab:HPO_params}{S2}\noindent\textbf{Table S2}

\end{document}